\documentclass[12pt]{iopart}      

\usepackage{graphicx}

\begin{document}

\title[PSI: A new method for the early detection of chaos]{The power spectrum
  indicator: A new, efficient method for the early detection of chaos.}

\author{Ch Vozikis$^1$, K Kleidis$^2$ and S Papaioannou$^1$}

\address{$^1$ Department of Civil Engineering
              and Surveying \&
              Geoinformatics Engineering,
              Technological Education
              Institute of Central Macedonia,
              621.24 Serres, Greece}
\address{$^2$  Department of Mechanical Engineering,
              Technological Education
              Institute of Central Macedonia,
              621.24 Serres, Greece}
\ead{\mailto{chriss@teicm.gr}, \mailto{kleidis@teicm.gr} and
  \mailto{pasta@teicm.gr}}

\begin{abstract}
To determine the  regular or  chaotic nature of the orbits in dynamical
 systems can be quite an issue. In  
this article, following Vozikis et al. (2000), we propose a new tool, namely,
the Power Spectrum Indicator (PSI), $\psi^2$, that enables us to determine, as
early as posible, whether an orbit of a two-dimensional map is chaotic or
not. This new method is based on the frequency analysis of a data series
constucted by recording the logarithm of the amplification factor of the
deviation vector of nearby orbits. Accordingly, two datasets are recorded and
the $\chi^2$-likelyhood of their power spectra is computed. Ordered orbits
have always the same power spectrum, so their $\chi^2 \equiv \psi^2$ acquires
a zero value. On the contrary, a chaotic orbit has a power spectrum that
varies with time, hence, chaotic orbits always exhibit a non-zero $\psi^2$
value. Even as regards {\it "sticky"} orbits, the PSI method is very effective
in the early detection of chaos, while the global behavior of the $\psi^2$
indicator can provide information (also) on the intense of the chaotic
behavior, i.e., on how {\it "strong"} or {\it "weak"} the associated chaos may
be.  
\end{abstract}

\noindent{\it keywords\/}: dynamics, chaos, chaotic orbits, detection of chaos.

\submitto{\jpa}

\section{Introduction}
A major issue in studying non-integrable dynamical systems is the
determination, as early as posible, of the orbits' (chaotic or not) nature. In
the pioneering work of H\'enon and Heiles (1964), when the related research
was limited to two-dimensional $(2D)$ systems, the study of the orbits' nature
was performed by means of the surface of section (SoS). When research extended
to more complex, three-dimensional $(3D)$ systems (where the method of SoS
cannot be applied), the problem was addressed in terms of the Lyapunov
characteristic numbers - LCN (Benettin et al. 1976, Froeschl\'e 1984). The LCN
method tracks the evolution of the deviation vector, $\vec{d}$, that connects
the positions of two nearby orbits in phase-space at each time-step or,
equivalently, at each iteration. Unfortunately both methods have the same
weakness: They cannot distinguish, early enough, a {\it "sticky"} chaotic
orbit from an ordered one (see, e.g., Contopoulos \& Voglis 1997).  

Since then, several other methods have been proposed. Some of them, are based
on the analysis of a time-series associated to the values of the generalized
coordinates or functions of these coordinates, as the rotation number method
(Contopoulos 1966), the frequency map analysis (Laskar et al. 1992, Laskar
1993) and the power spectrum of quasi-integrals analysis (Voyatzis \&
Ichtiaroglou 1992). Other methods, use the geodesic divergence of initially
nearby trajectories, as in the probability-density analysis of stretching
numbers (Froeschl\'e et al. 1993, Voglis \& Contopoulos 1994), the fast
Lyapunov indicators (Froeschl\'e et al. 1997), and the methods of alignment
indexes, namely, the small alignment indexes - SALI (Skokos 2001, Skokos et
al. 2003, 2004, Skokos \& Manos 2016) and the generalized alignment indexes -
GALI (Skokos et al. 2007, Skokos \& Manos 2016). Each and everyone of the
above methods has its own advantages and 
weaknesses; some of them are more efficient to addressing 2D systems rather
than their higher-dimensional counterparts, while others perform better on
mappings rather than flows. 

In this context, eightteen years ago, Vozikis et al. (2000) proposed a method
based on the frequency analysis (power spectrum) of stretching numbers. A
variant of this method was used also by Karanis \& Vozikis (2008). The method
is fast and efficient, but it has a major disadvantage. In order to 
decide whether an orbit is regular or not, one needs to visually inspect the
power spectrum and to classify it as either
representing a regular orbit or a chaotic one.  

In the present article, we are revisiting the method proposed by 
Vozikis et al. (2000),
introducing a major improvement, that may help us to override the previous
disadvantage. As a result of this new method, we end up with (just) a single
number that enables us to classify an orbit as being either regular or
chaotic.  

This article is organized as follows: In Section 2, we summarize the
power-spectrum method of Vozikis et al. (2000), in the context of which we
will set up also our new model. In Section 3, we perform a {\it
  goodness-of-fit}, $\chi^2$-analysis of several successive power spectra,
resulting from datasets that are associated with the deviation vectors of
three particular types of orbit, namely, regular, chaotic, and {\it
  "sticky"}. As a result of the aforementioned likelyhood analysis, a new
indicator of chaotic behavior is introduced (Section 4), which allows us to
classify any kind of orbit on a $2D$ mapping, as early as possible. Finally,
we conclude in Section 5.

\section{The model and the power spectrum method}

\subsection{The model}

One of the most frequently used test-models for studying chaotic motion is the
$2D$ {\it standard map}, appearing in literature in too many forms
(Lichtenberg \& Lieberman 1983, Ichikawa et al 1987, Aubry and Abramovici
1990, Contopoulos \& Voglis 1997, Gelfreich 1999, Lazutkin 2005 ). In the
present article, we adopt the following set of recursive  relations 
\numparts
\begin{eqnarray}
 J_{i+1} & = & J_i + k \cos(2 \theta_i)  ~~~ \mathrm{mod}(2 \pi) \: ,\\
 \theta_{i+1} & = & \theta_i + J_{i+1}  ~~~~~~~~~~   \mathrm{mod}(2 \pi) \:,
\label{eq:map} 
\end{eqnarray}
\endnumparts
where $J_i$ and $\theta_i$ are the associated action-angle variables, $i$
stands for the iteration number, and $k$ is the {\it "stochasticity
parameter"} (see, e.g., Lichtenberg \& Lieberman 1983). In what follows, we
consider that $k = 0.7$, a case where the standard map possess both regular
and chaotic regions. 
 
\begin{figure*}
\resizebox{\hsize}{!}
{\includegraphics*{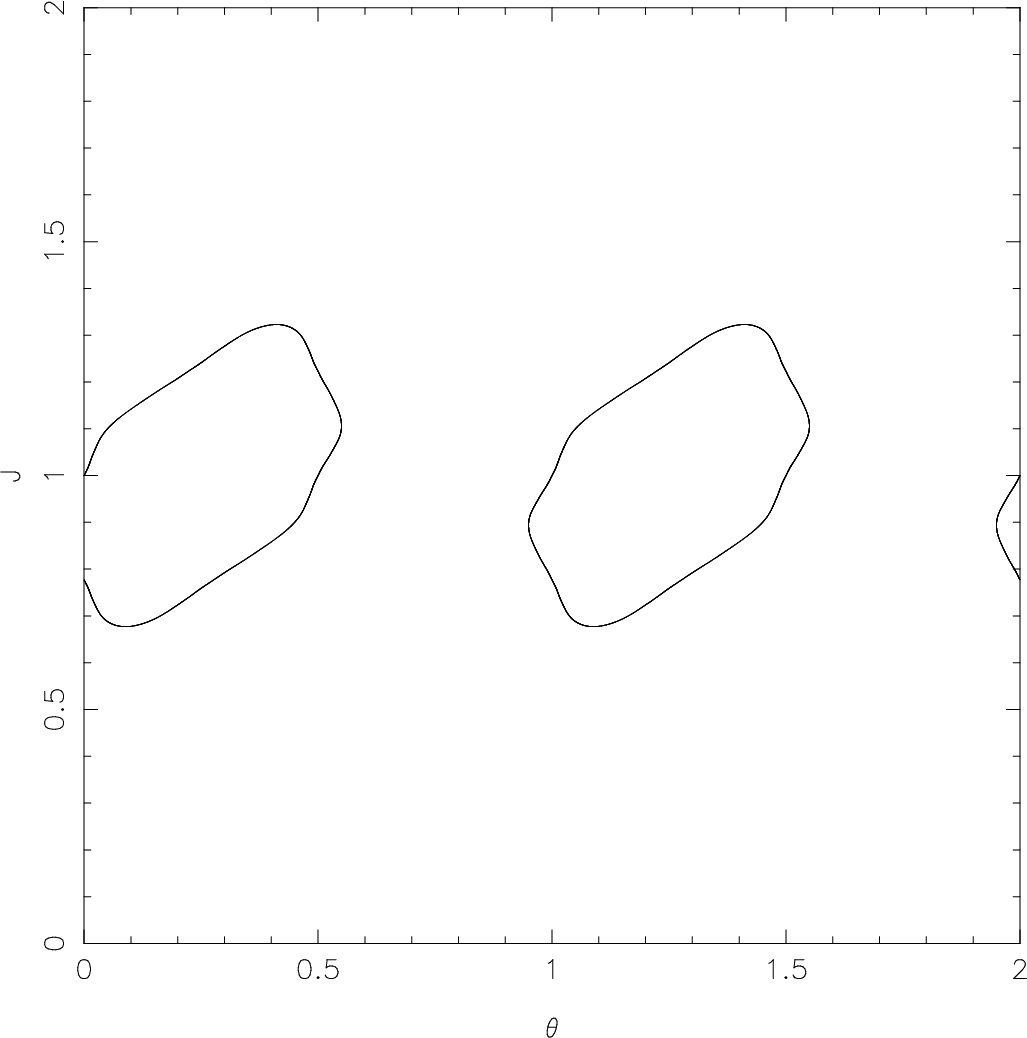}
\hspace{1cm}
\includegraphics*{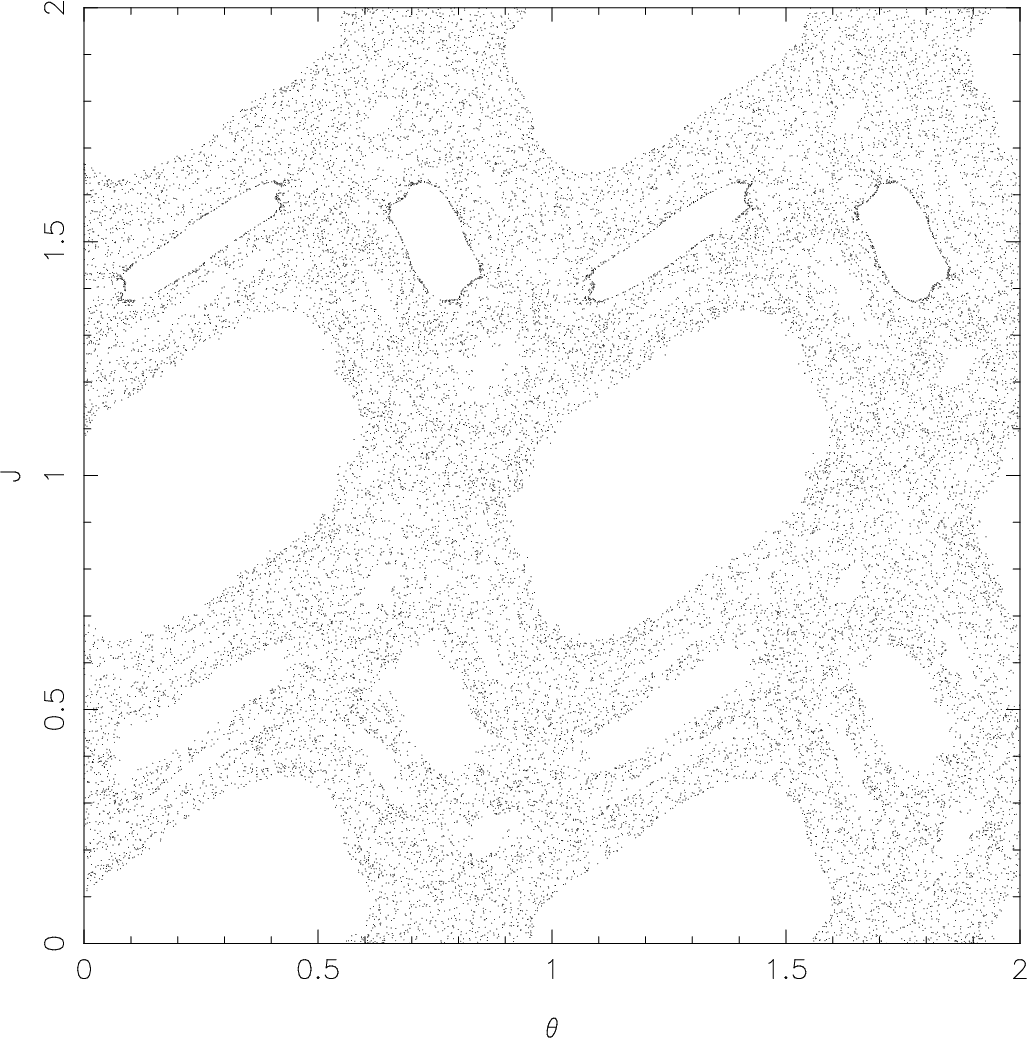}}
\caption{Successive points of two orbits on the {\it standard map}, when $k =
  0.7$. Left frame, an ordered orbit originating at $J_0 = \pi $,
  $\theta_0 = 1.5~\pi$; Right frame, a chaotic orbit originating at $J_0 =
  1.3~\pi $, $\theta_0 = 1.5~\pi$  (the axes units are in multiples of $\pi$). }
\label{fig:points1}
\end{figure*}

\begin{figure*}
\resizebox{\hsize}{!}
{\includegraphics*{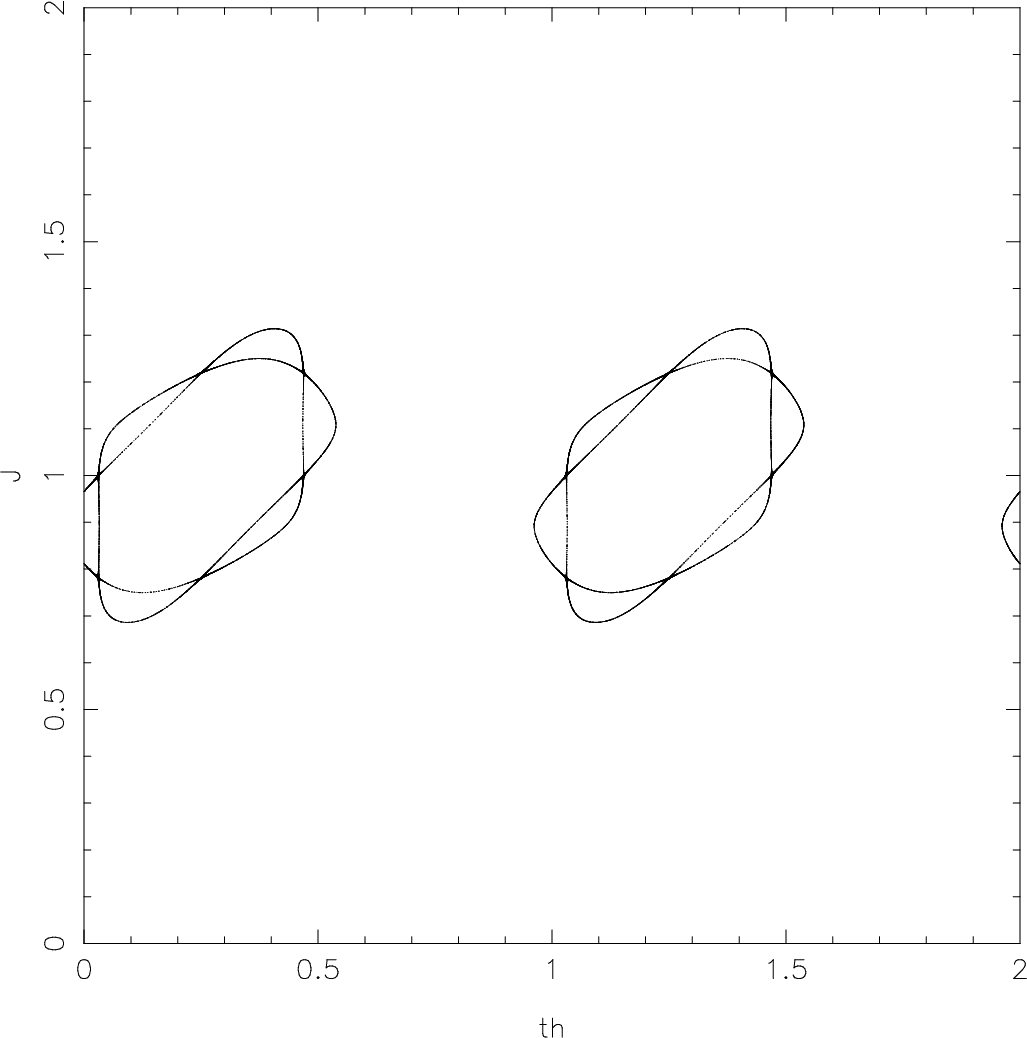}
\hspace{1cm}
\includegraphics*{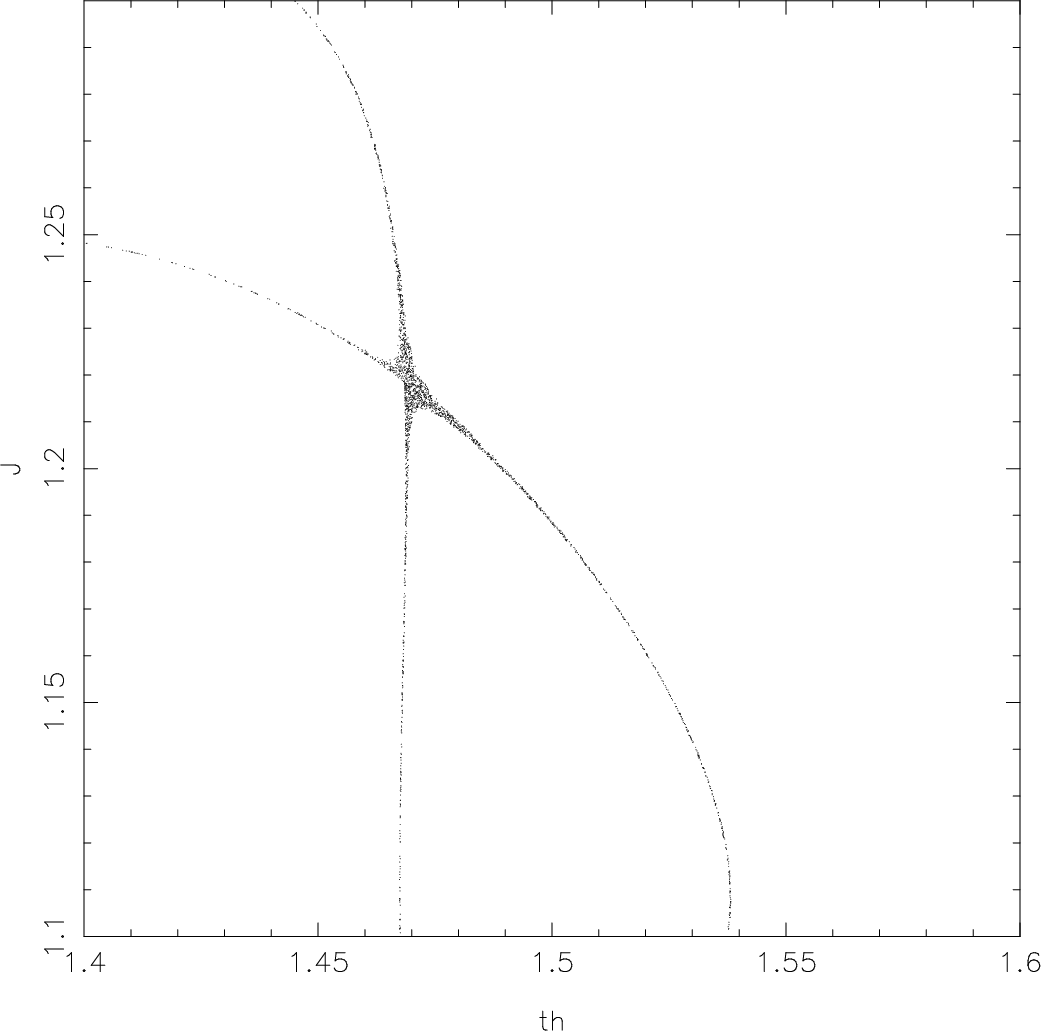}}
\caption{Successive points of a chaotic orbits on the {\it standard map}, when
  $k =   0.7$ originating at $J_0 = 1.1998 ~\pi $, $\theta_0 = 1.49~\pi$. Left
  frame, full view; Right frame, a zoom on the area near the separatrix of
  the orbit.} 
\label{fig:points2}
\end{figure*}

\begin{figure*}
\resizebox{\hsize}{!}
{\includegraphics*{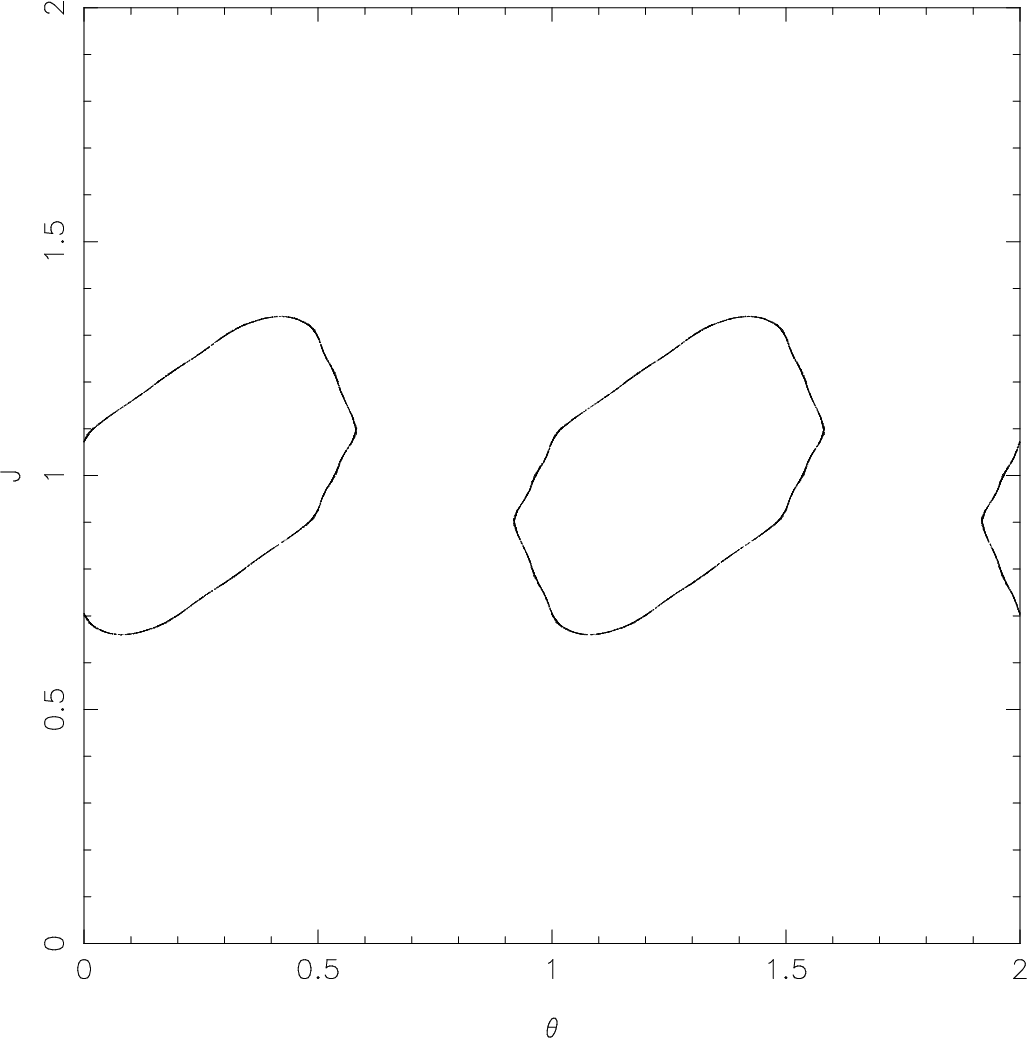}
\hspace{1cm}
\includegraphics*{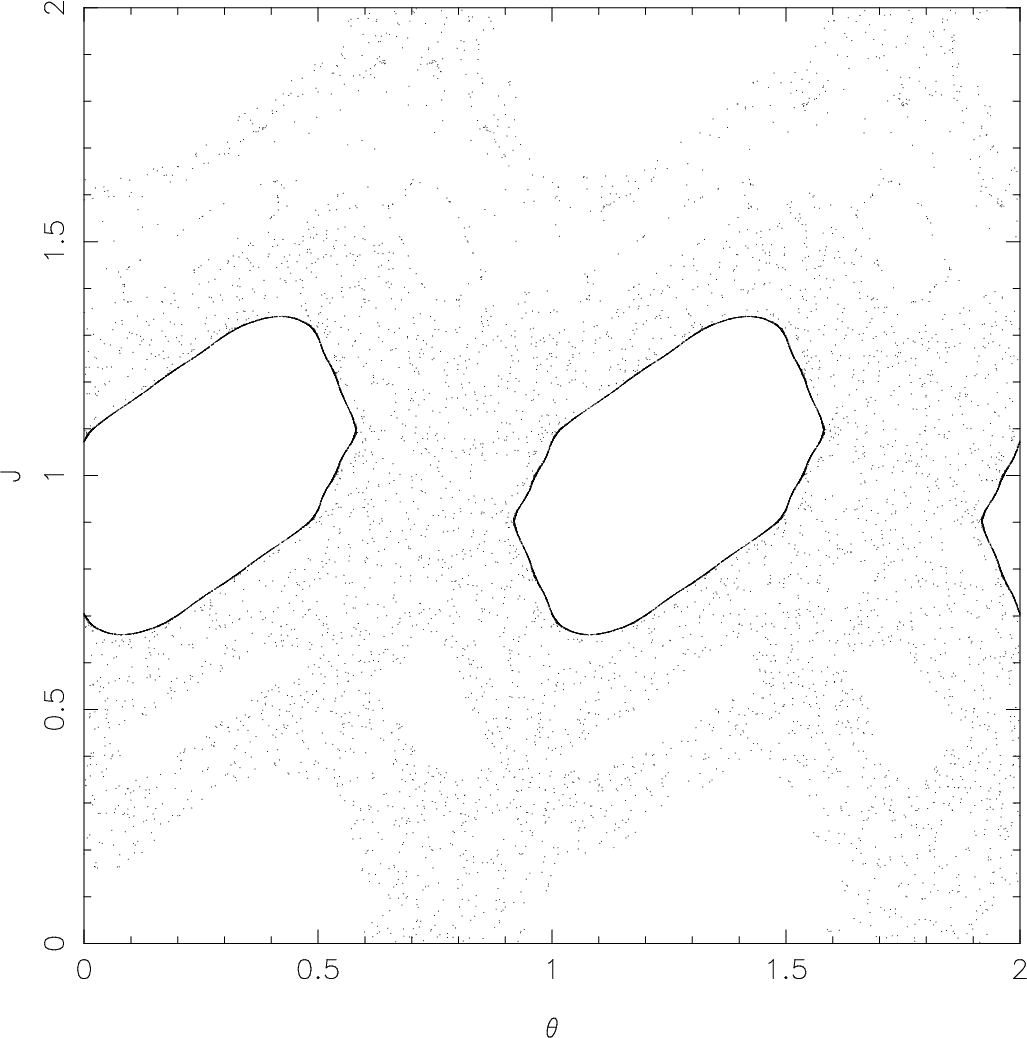}}
\caption{Successive points of a ``sticky'' orbit originating at 
$J_0 = \pi $, $\theta_0 = 1.538~\pi$. 
Left frame, the first 10\,000 points;
Right frame, the first 15\,000 points.} 
\label{fig:points3}
\end{figure*}

In figures \ref{fig:points1}, \ref{fig:points2} and \ref{fig:points3}
 we present four characteristic orbits of the
standard map, for $k=0.7$. Notice that, in all figures, $J_i$ and $\theta_i$ 
are given as multiples of $\pi$. The two frames of figure \ref{fig:points1}
represent the first 20\,000 points of an ordered orbit, with initial
conditions $J_0 = \pi $, $\theta_0 = 1.5~\pi$ (left frame), and of a chaotic
orbit, with initial position at $J_0 = 1.3~\pi $, $\theta_0 = 1.5~\pi$ (right
frame). The left frame of figure \ref{fig:points2} represents a chaotic orbit,
originating at $J_0 
= 1.1998~\pi $, $\theta_0 = 1.49~\pi$. Although the orbit looks ordered, a
zoom (right frame) on the area near the separatrix reveals its chaotic
nature. Finally the two frames of figure \ref{fig:points3} represent a {\it
  "sticky"} orbit   
with initial position $J_0 = \pi $, $\theta_0 = 1.538~\pi$. The left frame
shows the first 10\,000 successive points, while the right one presents the
corresponding 15\,000 ones. This {\it "sticky"} orbit, although it is chaotic,
behaves, at least macroscopically, like an ordered one for about 11\,000 
iterations.  

\subsection{The power spectrum (PSOD) method}

In order to decide on the nature of an orbit (chaotic or not), Vozikis et
al. (2000) proposed the Power Spectrum of Orbits Divergence (PSOD)
method. This method consists in the numerical integration of the orbit
originating at $\left( J_0, \theta_0 \right)$, along with a nearby one,
originating at an infinitesimally-close distance in phase space,
$\vec{d_0}=\left( dJ_0, d\theta_0 \right)$, i.e., of an initial position
$J_0^{\prime} = J_0 + dJ_0$, $\theta_0^{\prime} = \theta_0 + d\theta_0$. At
each iteration, $i$, the quantity  
\begin{equation}
q_i = \ln (d_i/d_0) \: ,
\label{eq:q}
\end{equation}
where $\vec{d_i}=\left ( dJ_i, d\theta_i \right )$, is recorded. To calculate
$\vec{d_i}$, instead of integrating numerically also the second, nearby,
orbit, one can use the so called {\it variational equations}  
\begin{eqnarray}
 dJ_{i+1} & = & dJ_i - 2k \sin (2 \theta_i) d \theta_i \nonumber \\ d \theta_{i+1} & = & d \theta_i + dJ_{i+1} .
\label{eq:var_map}
\end{eqnarray}
These equations can be easily obtained, by
substituting in the standard map $J^{\prime} = J + dJ$ and $\theta^{\prime} =
\theta + d \theta$, and expanding $\sin (2 \theta^{\prime})$ as a Taylor
series, keeping only first order terms in $dJ$ and $d \theta$. Prior to any
iteration, the deviation vector, $\vec{d}$, is renormalized, upon
multiplication by the factor $d_0/d_i$. In other words, at each and every $i$,
although $\vec{d_i}$ retains the orientation acquired, its norm remains equal
to $d_0$. Thus, after tracking this orbit for $N$ successive iterations, we
are left with a series of consecutive $q_i$ $( i = 1, 2, ..., N)$.   

The power spectrum of the aforementioned $q$-series can be obtained by taking
the discrete Fourier transform of $q_k$, multiplied by a {\it window
  function}, $w_k$,  
\begin{equation}
Q_j = \sum_{k=0}^{2N-1} q_k w_k e^{2 \pi i \frac{j}{N} k} ~~~~~j = 0,..., \left( 2 N - 1 \right) \: .
\end{equation}
As window function we use the so-called {\it Hanning window} (see, e.g., Press
et al. 1992). In this context, the power spectrum, $P(f_j)$, is defined over a
set of $M = N+1$ frequencies, $f_j$, as  
\begin{eqnarray}
\label{eq:Power}
P(f_0) & = & \frac{1}{W} ~ \vert Q_0 \vert^2 \: , \nonumber \\ 
P(f_j) & = & \frac{1}{W} ~ \left( \vert Q_j \vert^2 + \vert Q_{2N-j} \vert^2 \right )~~~~~ j = 1,..., \left ( N-1 \right ) \: , \\
P(f_c) & = & \frac{1}{W}~ \vert Q_{N} \vert^2 \: , \nonumber  
\end{eqnarray}
where we have set 
\begin{equation}
W = 2N \sum_{k=0}^{2N-1}w_k^2 \: . 
\end{equation}
In Eqs. (\ref{eq:Power}) , the frequency $f_c = f_{N}$ is the Nyquist
frequency, which, in 
the case of the standard map, is 
\begin{equation}
f_c = \frac{1}{2}.
\end{equation}
The group of frequencies embraced by the power spectrum of
Eqs. (\ref{eq:Power}) , is given by:  
\begin{equation}
f_j = f_c~\frac{j}{M}~~~~~~~j = 0, ..., M
\label{eq_freqs}
\end{equation}
More elaborated details on the calculation of the power spectrum can be found
in the book {\it "Numerical Recipes"} by Press et al. (2001). In this article,
as far as the computation of the power spectrum is concerned, we use two sets
of successive $2 \times N$ data, which overlap with each other by one half of
their lengths. In other words, the whole dataset $q_i$ involved in a single
calculation of the power spectrum is $N_S = 3 \times N$, where $N$ is a power
of 2.  

\begin{figure*}
\resizebox{\hsize}{!}
{\rotatebox{270}{\includegraphics*{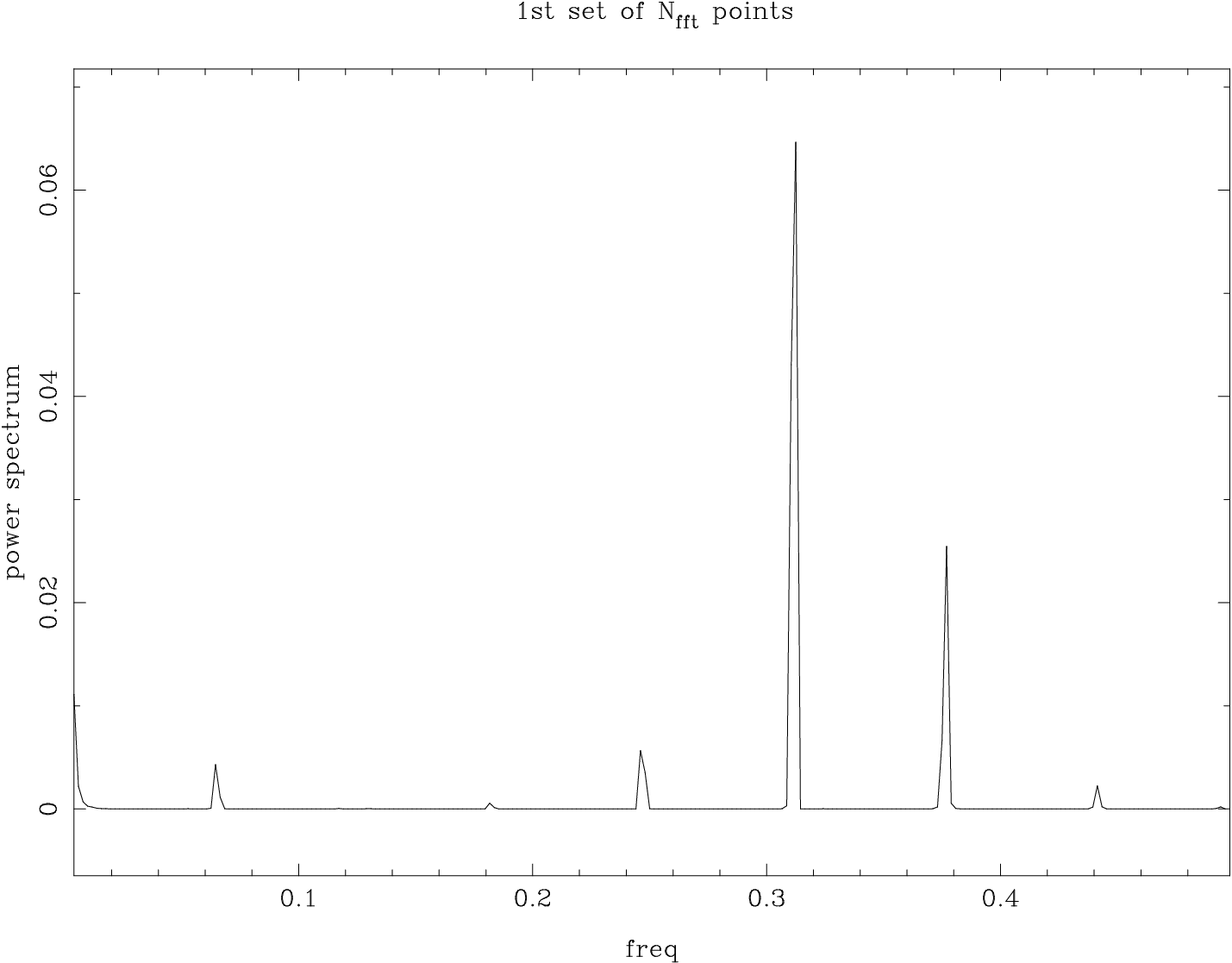}}
\hspace{1cm}
\rotatebox{270}{\includegraphics*{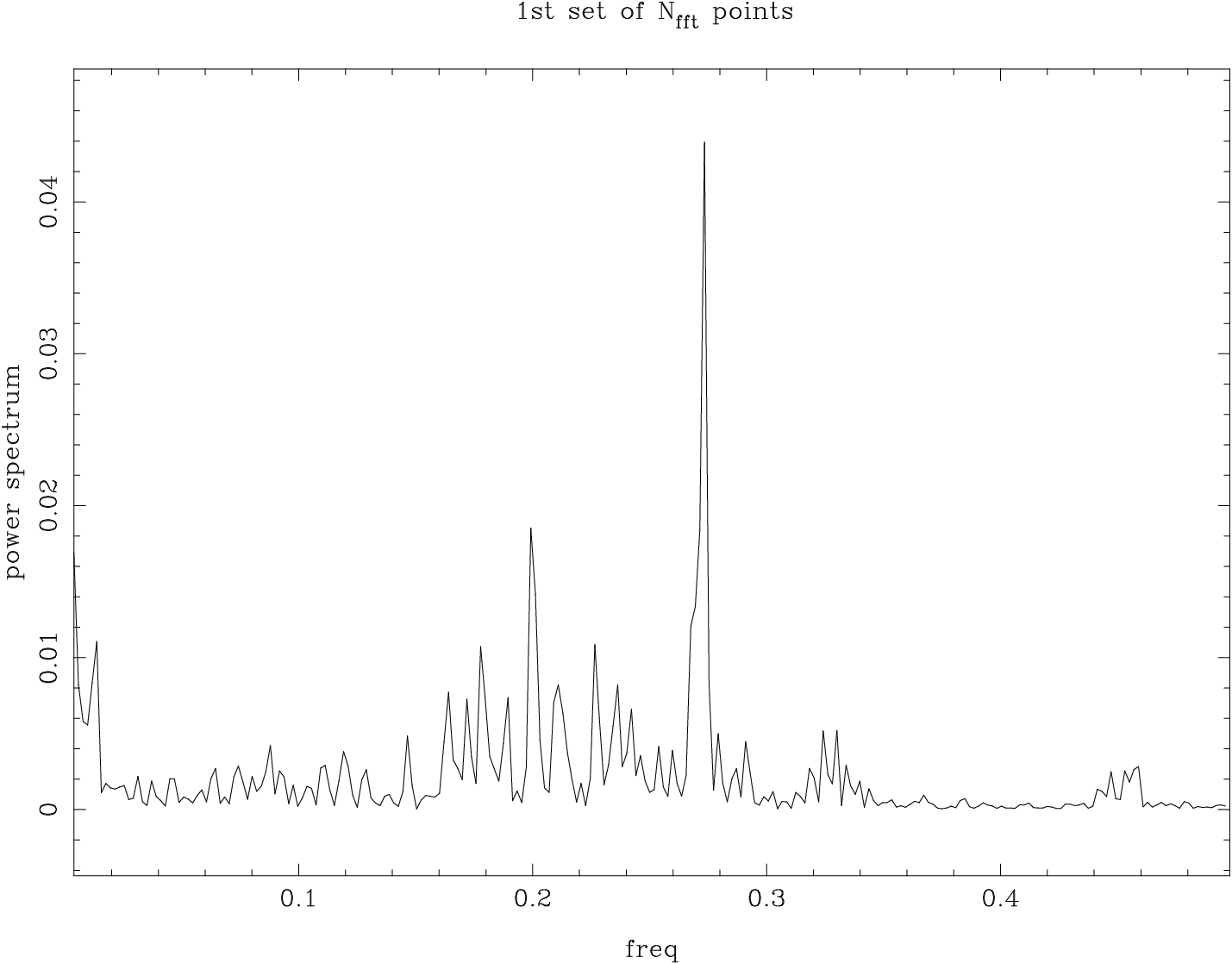}}}
\caption{The power spectra of two orbits for $N_s = 3 \times 256$
  iterations. Left frame: The power spectrum associated to the ordered orbit
  originating at $J_0 = \pi $, $\theta_0 = 1.5~\pi$. Right frame: The
  corresponding quantity as regards the chaotic orbit originating at $J_0 =
  1.3 \pi $, $\theta_0 = 1.5~\pi$.} 
\label{fig:psod_oc}
\end{figure*}

Figure \ref{fig:psod_oc} presents the power spectra associated with the two
orbits of figure \ref{fig:points1}. The spectrum on the left
frame is for the regular orbit, while the one on the right frame is for the
chaotic. One may easily decide on the nature of a particular orbit, simply by
inspecting its spectrum. Ordered motion corresponds to a power spectrum with
only a few spikes in certain frequencies. On the contrary, a chaotic orbit
exhibits a spectrum that consists of almost all frequencies, with varying
amplitudes. 

\section{A novel method}

\subsection{The basic idea}

The basic idea behind our new method is that, ordered motion is a kind of
quasi-periodic motion. Thus, the power spectrum of an ordered orbit will
always have some sort of characteristic frequencies, i.e., it will be
independent of time (or, equivalently, of the iteration number). On the
contrary, chaotic motion is a sort of random walk. Hence, we
expect that for a chaotic orbit, two different sets of $q_i$ (eq. \ref{eq:q})
recorded on different times (iterations) will be completely different,
since the motions posses no periodicity at all. 

\begin{figure*}
\resizebox{\hsize}{!}
{\rotatebox{270}{\includegraphics*{fig02l.ps}}
\hspace{1cm}
\rotatebox{270}{\includegraphics*{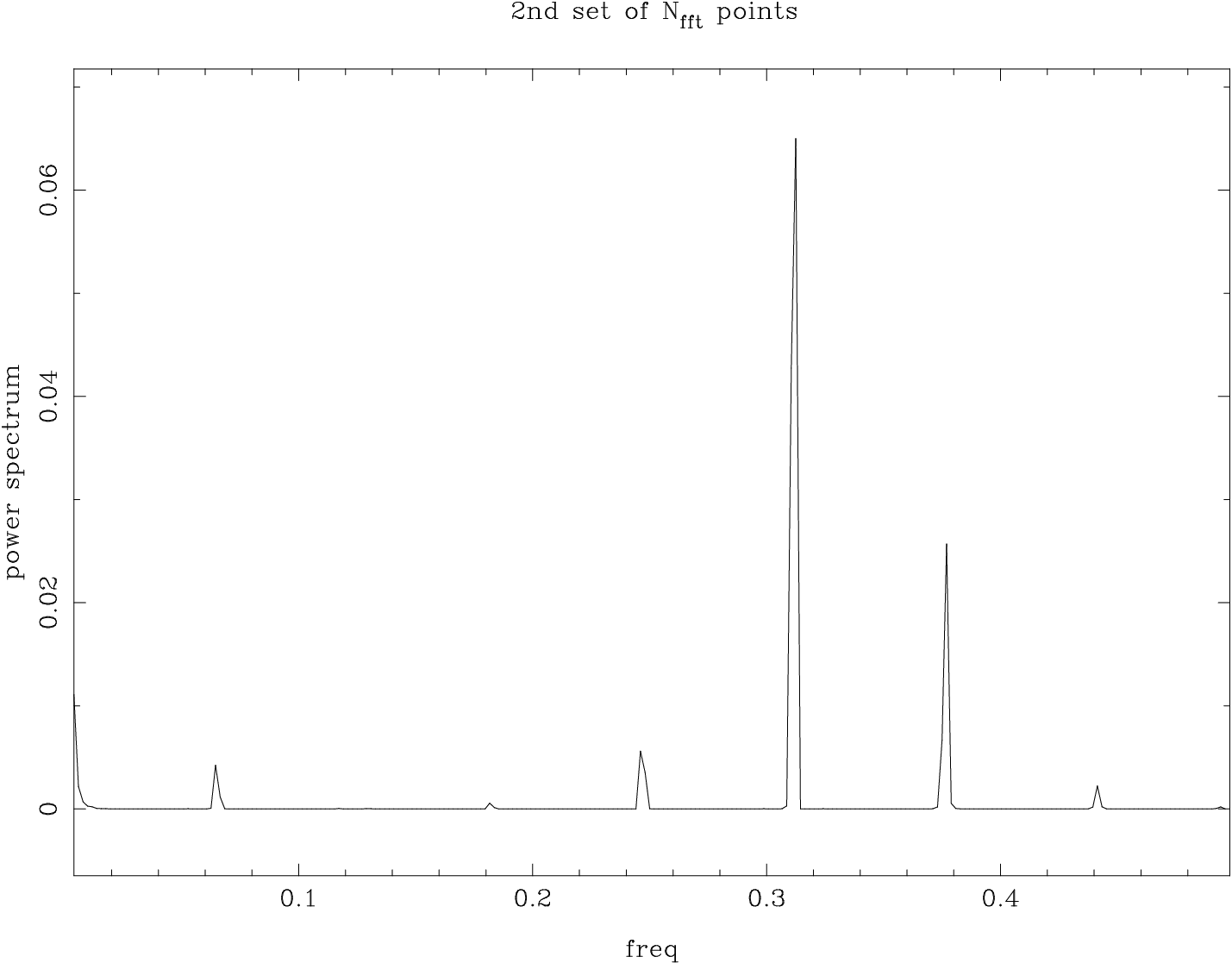}}}
\caption{The PSODs of the first $N_S = 3 \times 256 = 768$ iterations (left)
  and the next $N_S$ iterations (right) of an ordered orbit originating at
  $J_0 = \pi $, $\theta_0 = 1.5~\pi$.} 
\label{fig:comp_psd_o}
\end{figure*}

\begin{figure*}
\resizebox{\hsize}{!}
{\rotatebox{270}{\includegraphics*{fig03l.ps}}
\hspace{1cm}
\rotatebox{270}{\includegraphics*{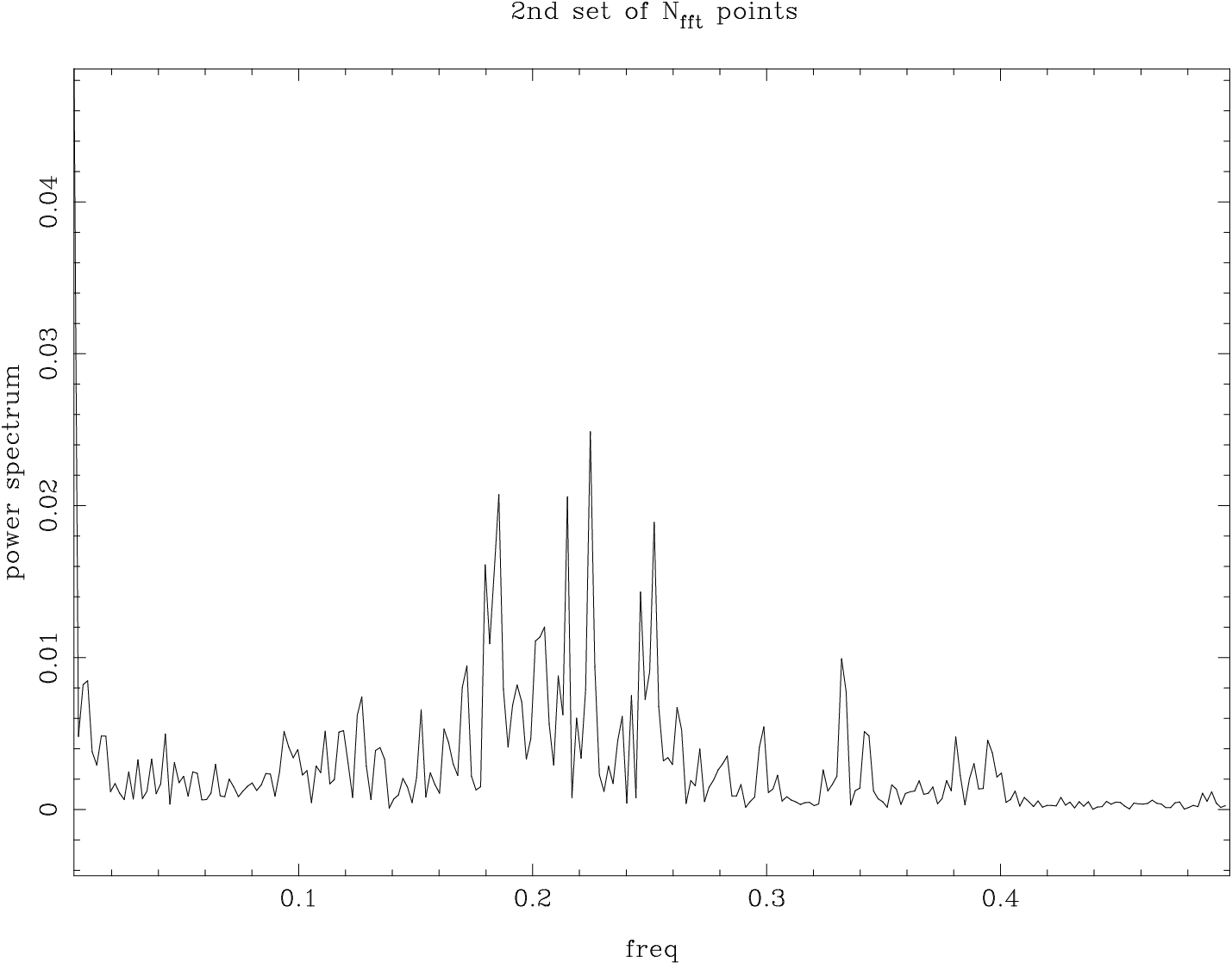}}}
\caption{Same as figure \ref{fig:comp_psd_o}, but, this time, for a chaotic
  orbit originating at $J_0 = 1.3~\pi $, $\theta_0 = 1.5~\pi$.} 
\label{fig:comp_psd_c}
\end{figure*}

\begin{figure*}
\resizebox{\hsize}{!}
{\rotatebox{270}{\includegraphics*{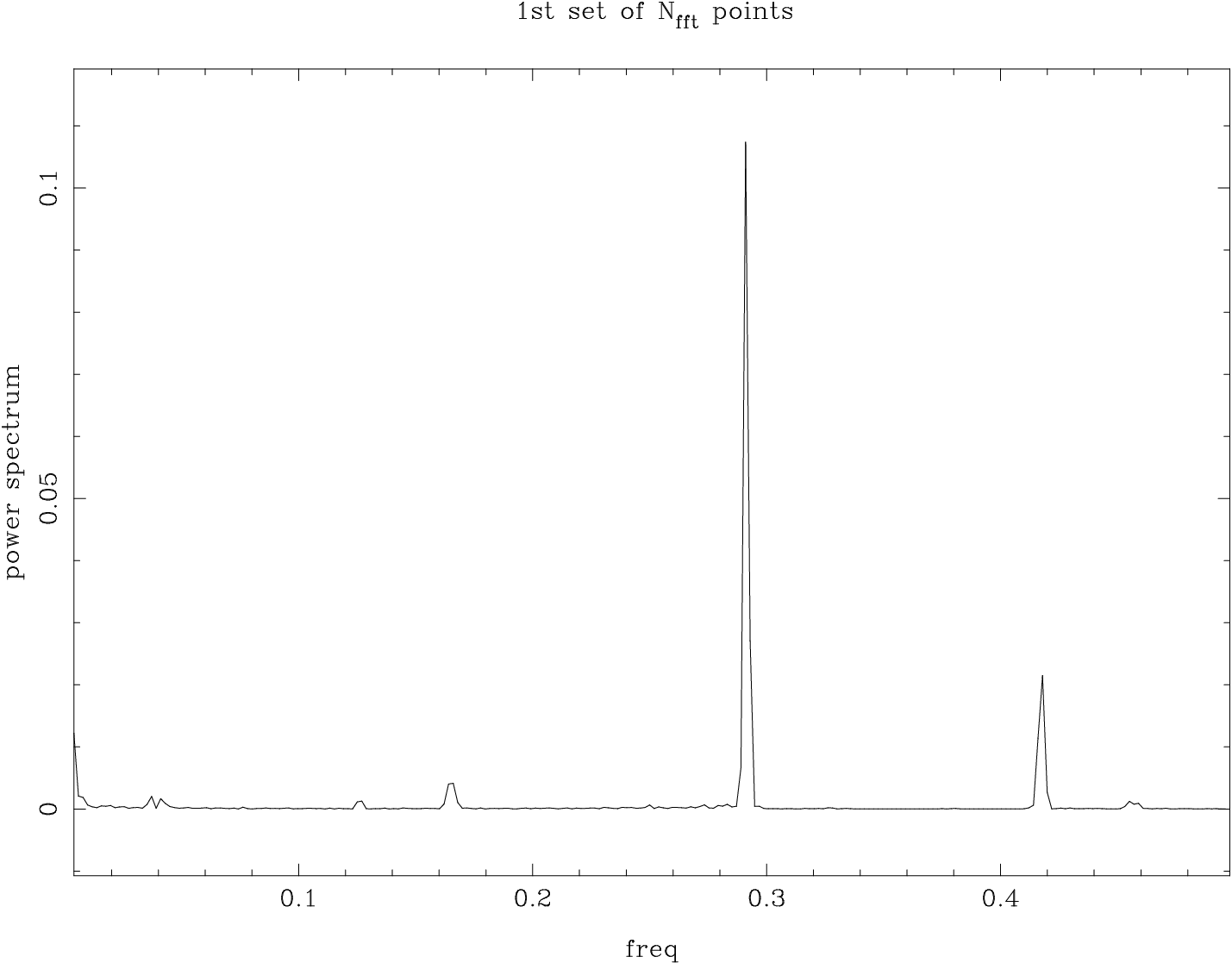}}
\hspace{1cm}
\rotatebox{270}{\includegraphics*{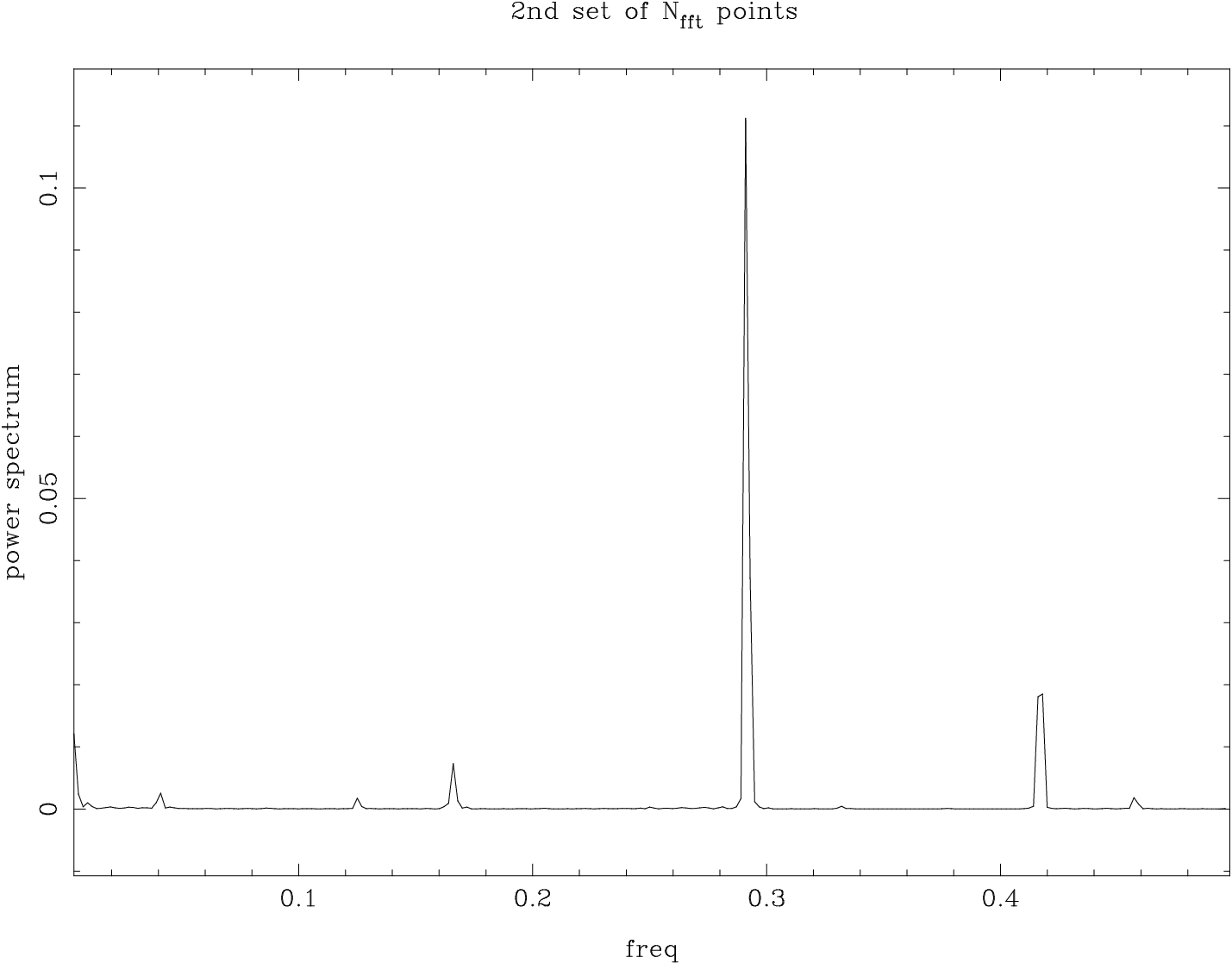}}}
\caption{Same as figures \ref{fig:comp_psd_o} and \ref{fig:comp_psd_c}, but
  now, as far as the {\it "sticky"} orbit originating at $J_0 = \pi $,
  $\theta_0 = 1.538~\pi$ is concerned.} 
\label{fig:comp_psd_s}
\end{figure*}

The validy of this idea can be easily inspected in figures
\ref{fig:comp_psd_o} to \ref{fig:comp_psd_s}. In all three figures, at first,
we follow a specific orbit for $N_{S} = 3 \times 256$ iterations. The spectrum
of the associated set of $q_i$ is presented on the left frame of each
figure. Next, we follow the same orbit for another $N_{S}$ iterations, thus
creating a second set of $q_i$. The spectrum of this second set is presented
on the right frame. In figure \ref{fig:comp_psd_o}, the two successive spectra
of an ordered orbit are presented. These two spectra appear to be the same. On
the contrary, as far as a chaotic orbit is concerned, the difference between
the associated two spectra is more than obvious (cf. figure
\ref{fig:comp_psd_c}). 

Detecting the chaotic nature of a {\it "sticky"} orbit is the most challenging
case, since such an orbit can behave like an ordered one for many
iterations. A reliable {\it chaos-detecting tool} must reveal the true
identity of these orbits as early as possible. Figure \ref{fig:comp_psd_s}
shows the two spectra of a {\it "sticky"} orbit. We see that these spectra
resemble the ones of an ordered orbit. However, although they appear to be the
same, upon a closer look we see that they are not: There are small differences
at those frequencies that are associated to low amplitudes.

\subsection{On the $\chi^2$-likelyhood of the PSOD method}

In order to see if and how much the PSOD of the same orbit
changes we perform a $\chi^2$-statistics, as regards two PSODs of a
 particular orbit. 

Let $P_j$ be 
the power spectrum $P \left ( f_j \right )$ of the first PSOD $(j = 1, 2,
... , N_S)$ and $P_j^{\prime}$ the corresponding spectrum, $P \left (
f_{j^{\prime}} \right )$, of the second PSOD $(j^{\prime} = N_S + 1, ... ,
i)$. Their $\chi^2$ likelyhood is, then, defined as (see, e.g., Chapter 14.3
of Press et al. 1992) 
\begin{equation}
\chi^2 = {\sum_{j=0}^M {\frac{\left ( P_j^{\prime} - P_j \right )^2}{P_j^{\prime} + P_j}}} \: .
\label{eq:chi2}
\end{equation}
Since the values of $P_j^{\prime}$ of the second PSOD depend on the particular
set of $q_i$, we consider that the second PSOD, $P_j^{\prime}$, is a function
of the iteration number $i$, namely,     
\begin{eqnarray}
P_j & = & {\rm PSOD} \left[ q \left ( 1 \right ),~...,~q(N_S) \right ] \: , \nonumber \\
P_j^{\prime} (i) & = & {\rm PSOD} \left [ q \left ( i-N_S+1 \right ),~...,~q(i) \right ] \: .
\label{eq:Pj}
\end{eqnarray}
As a consequense, now, $\chi^2$ becomes also a function of the iteration number, $i$.
\begin{equation}
\chi^2 = \chi^2(i) = {\sum_{j=0}^M{\frac{\left ( P_j^{\prime} (i) - P_j \right )^2}{P_j^{\prime} (i) + P_j}}}. 
\label{eq:chi2i}
\end{equation}

\begin{figure*}
\resizebox{\hsize}{!}
{\rotatebox{270}{\includegraphics*{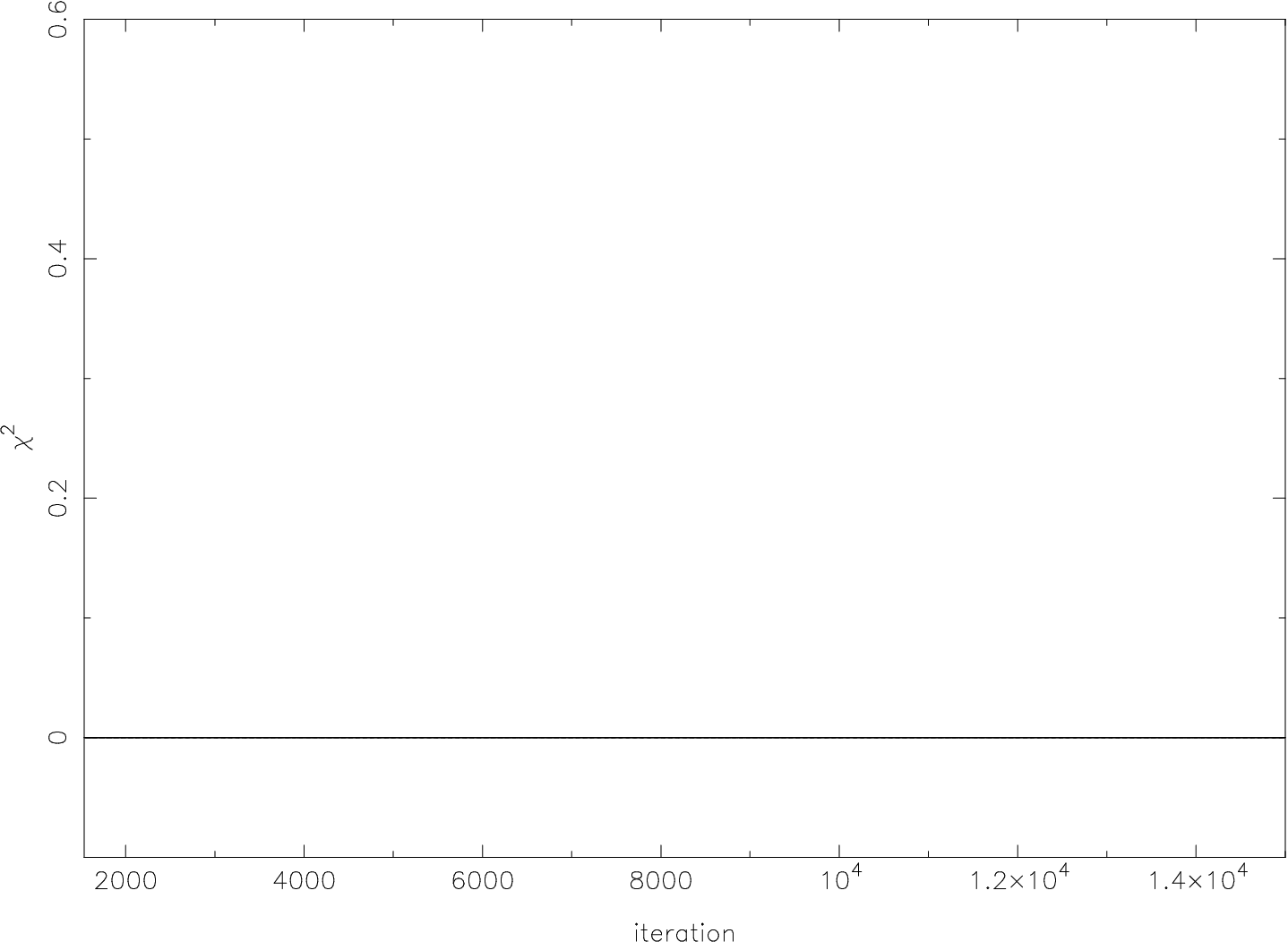}}
\hspace{1cm}
\rotatebox{270}{\includegraphics*{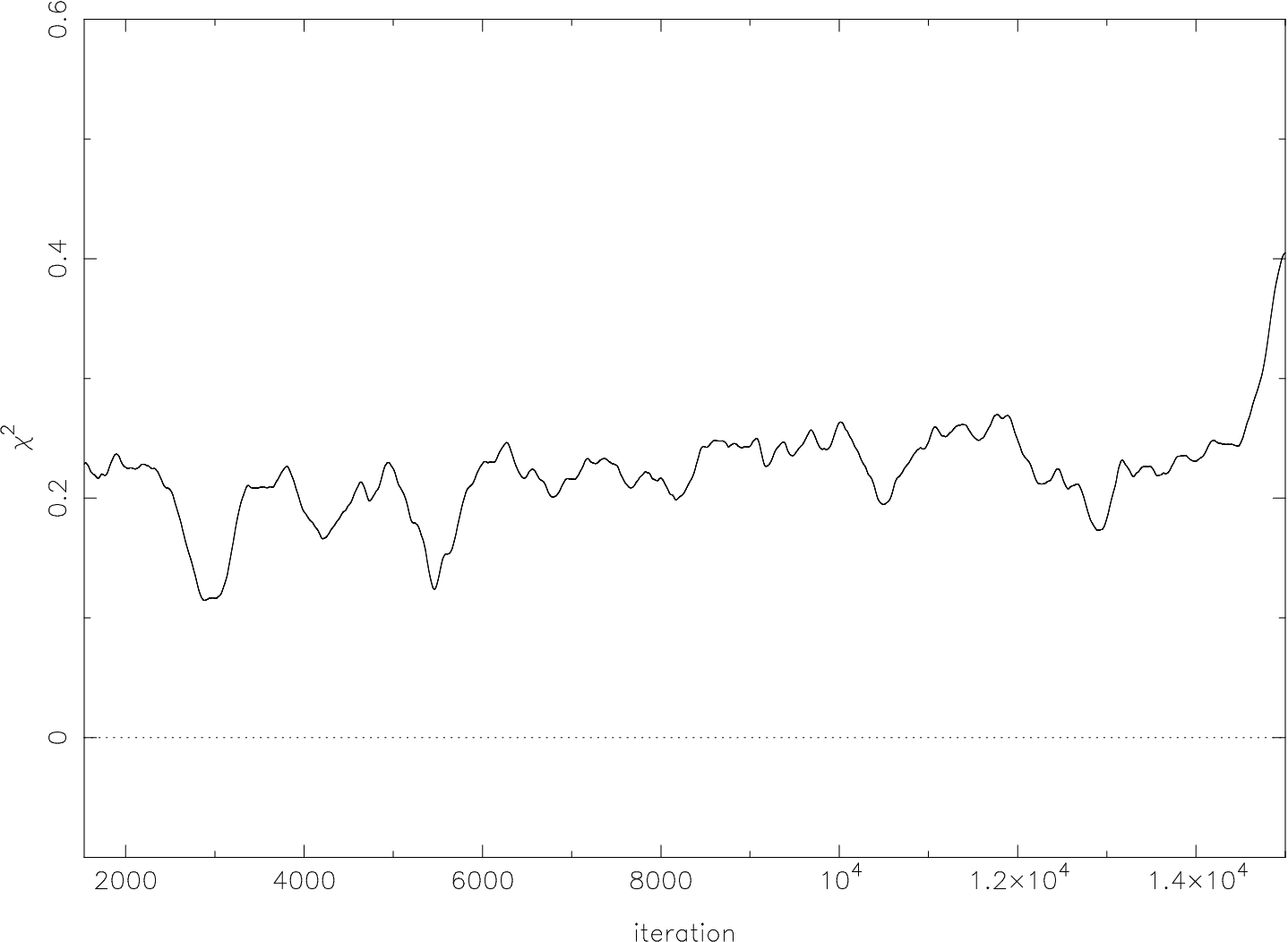}}}
\vskip 0.1cm
\resizebox{\hsize}{!}
{\rotatebox{270}{\includegraphics*{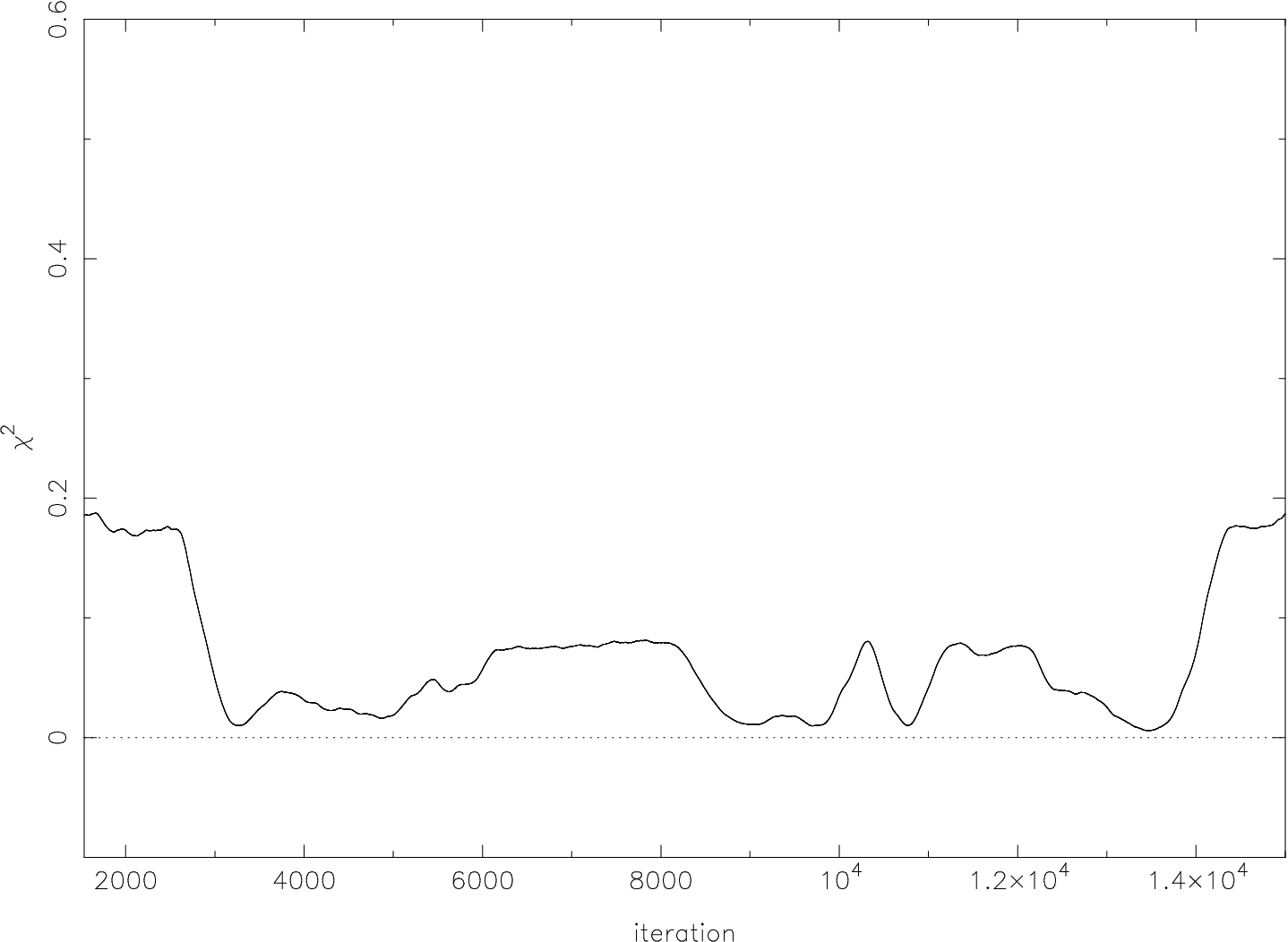}}
\hspace{1cm}
\rotatebox{270}{\includegraphics*{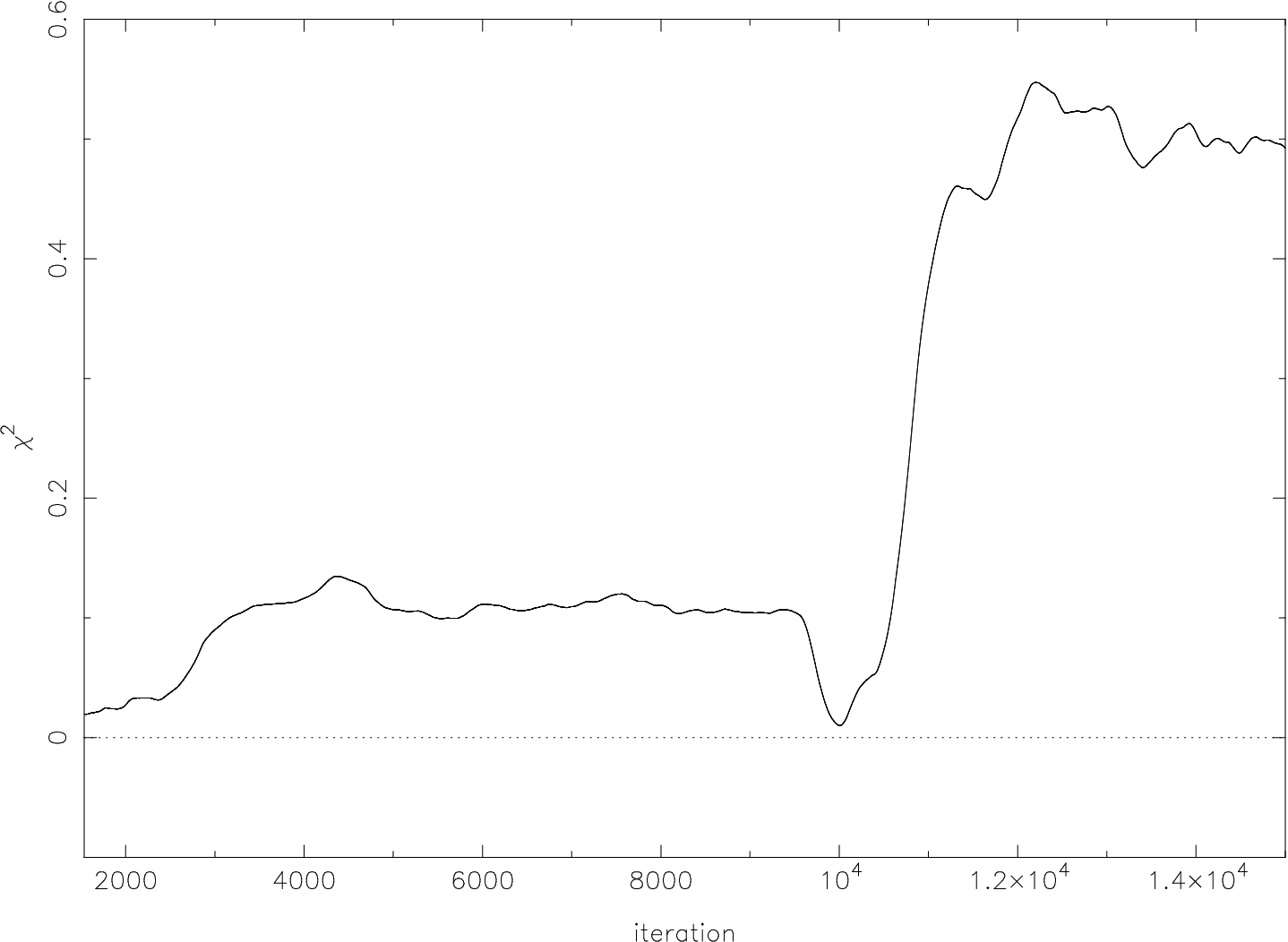}}}
\caption{$\chi^2(i)$-likelyhood of the PSODs corresponding to the orbits of
  figure 1. Top left frame, the ordered orbit originating at $J_0 = \pi $,
  $\theta_0 = 1.5~\pi$; Top right frame, the chaotic orbit originating at $J_0
  = 1.3~\pi$, $\theta_0 = 1.5~\pi$; Botton left frame, the second chaotic
  orbit originating at $J_0 = 1.1998~\pi $, $\theta_0 = 1.49~\pi$; Bottom
  right frame, the {\it "sticky"} orbit originating at $J_0 = ~\pi $,
  $\theta_0 = 1.538~\pi$.}  
\label{fig:chi2}
\end{figure*}

To begin with, let us examine how $\chi^2(i)$ behaves for various types of
orbits (i.e., ordered, chaotic, or {\it "sticky"}) and if the idea suggested
in Section 3.1 is valid, in the sense that it can give us reliable results as
far as the early prediction of chaos is concerned. Figure \ref{fig:chi2} shows
the evolution of $\chi^2(i)$ of the PSODs as a function of the iteration
number, $i$, for the four orbits of figures \ref{fig:points1},\ref{fig:points2}
and  \ref{fig:points3}. The top left
frame corresponds to the ordered orbit, having initial conditions 
$J_0 = \pi$, $\theta_0 = 1.5~\pi$. As we can see, the corresponding
$\chi^2(i)$ value 
remains zero, for every $i$. As we have already discussed in Section 3.1, an
ordered orbit represents a sort of quasi-periodic motion and, thus, its PSOD
will exhibit only a few, characteristic frequencies, remaining time-invariant
(or, equivalently, invariant with respect to the iteration number, $i$). On
the contrary, chaotic orbits behave in a definitely non-periodic manner and,
therefore, their PSOD will be completely different at different times
(equivalently, at different values of the iteration number, $i$). The top
right frame of figure \ref{fig:chi2} depicts the evolution of $\chi^2(i)$ for
a chaotic orbit, originating at $J_0 = 1.3~\pi$, $\theta_0 = 1.5 ~ \pi$. As we
can see, in this case, $\chi^2(i)$ starts from a value of $0.23$ and
fluctuates, but it never gets a zero value. A similar behavior can be seen
also in the evolution of $\chi^2(i)$ of the second chaotic orbit, originating
at $J_0 = 1.1998~\pi $, $\theta_0 = 1.49~\pi$. This orbit, although it
presents very week chaos (cf. figure \ref{fig:points2}),
exhibits, from the very beginning, a value of $\chi^2(i)$ around $0.18$. The
final, bottom right frame of figure \ref{fig:chi2} represents the evolution of
$\chi^2(i)$ associated to the {\it "sticky"} orbit, originating at $J_0 = \pi
$, $\theta_0 = 1.538 ~ \pi$. We can see that, even in the beginning, where it
is not easy at all to 
(visually) distinguish the difference between the two PSODs (cf. figure
\ref{fig:comp_psd_s}), the $\chi^2(i)$ method can reveal the true nature of
the orbit, acquiring a value of $0.03$. To put it more clearly, in this case,
the non-zero value of $\chi^2(i)$ of the two PSODs indicates that there are
differencies between them and, therefore, in the end, the orbit will exhibit a
clearly chaotic behavior. Indeed, as the time (or, equivalently, the iteration
number, $i$) goes by, $\chi^2(i)$ increases and, when the particular orbit
leaves the {\it "sticky"} region and enters into the {\it "big chaotic sea"},
it climbs to values higher than $0.5$! 

\section{The power spectrum indicator, $\psi^2$}

So far, we have seen that, as regards two PSODs of a particular orbit, the
$\chi^2(i)$-likelyhood analysis can give us important information on whether
this orbit is ordered or not. Clearly, a non zero value of $\chi^2$ suggests
that the orbit is definitely chaotic. In this context, the results of Section
3.2 may give rise to the following questions: 

\begin{enumerate}

\item Why does the $\chi^2$ value corresponding to the {\it "sticky"} orbit
  rises to such high values, as compeared to the chaotic orbits presented in
  figure \ref{fig:chi2}? 

\item Can we modify the method, in a way that it can give us (also) a clear
  indication on the degree of chaos? 

\end{enumerate}

\begin{figure*}
\resizebox{\hsize}{!}
{\rotatebox{270}{\includegraphics*{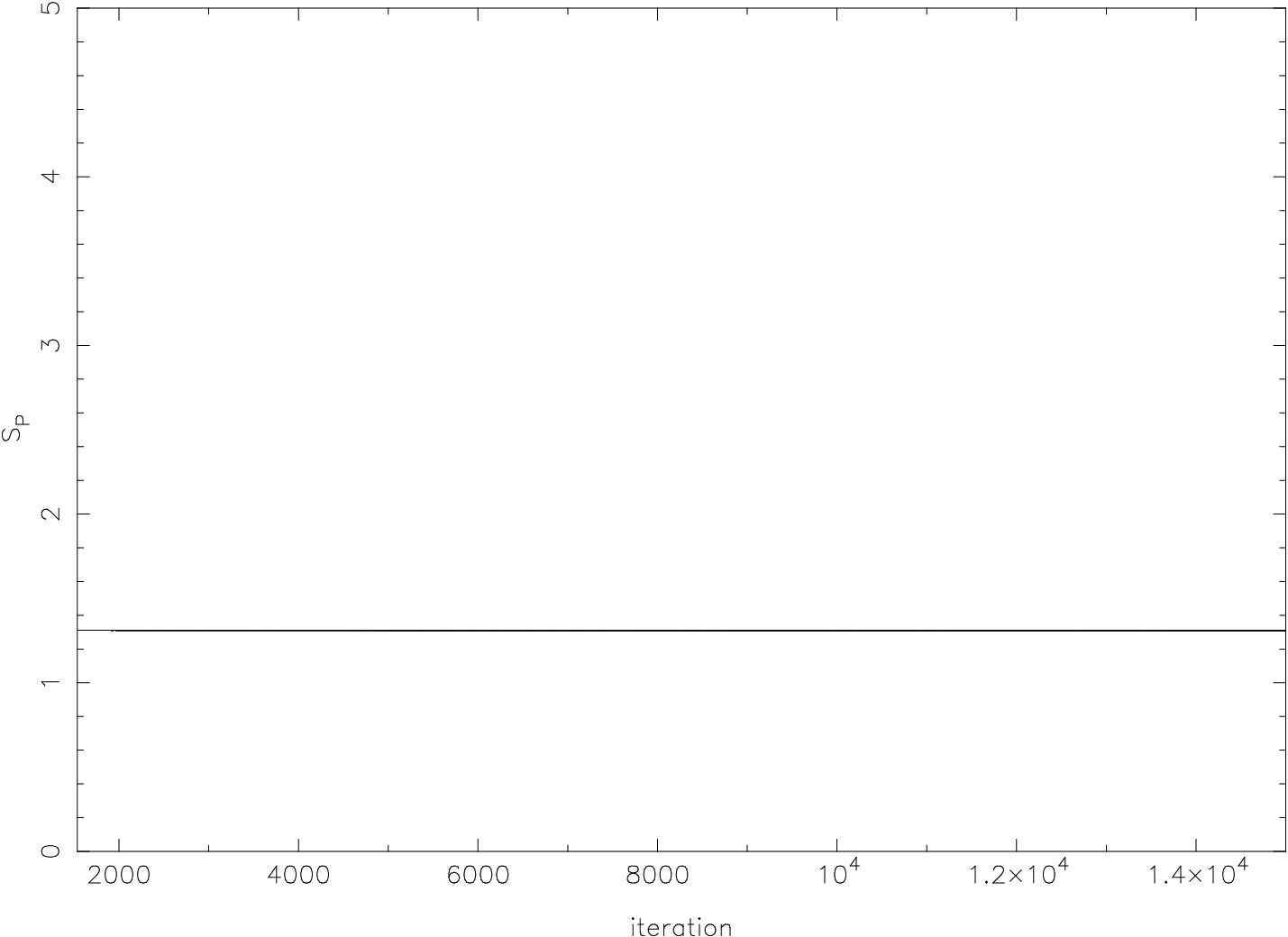}}
\hspace{1cm}
\rotatebox{270}{\includegraphics*{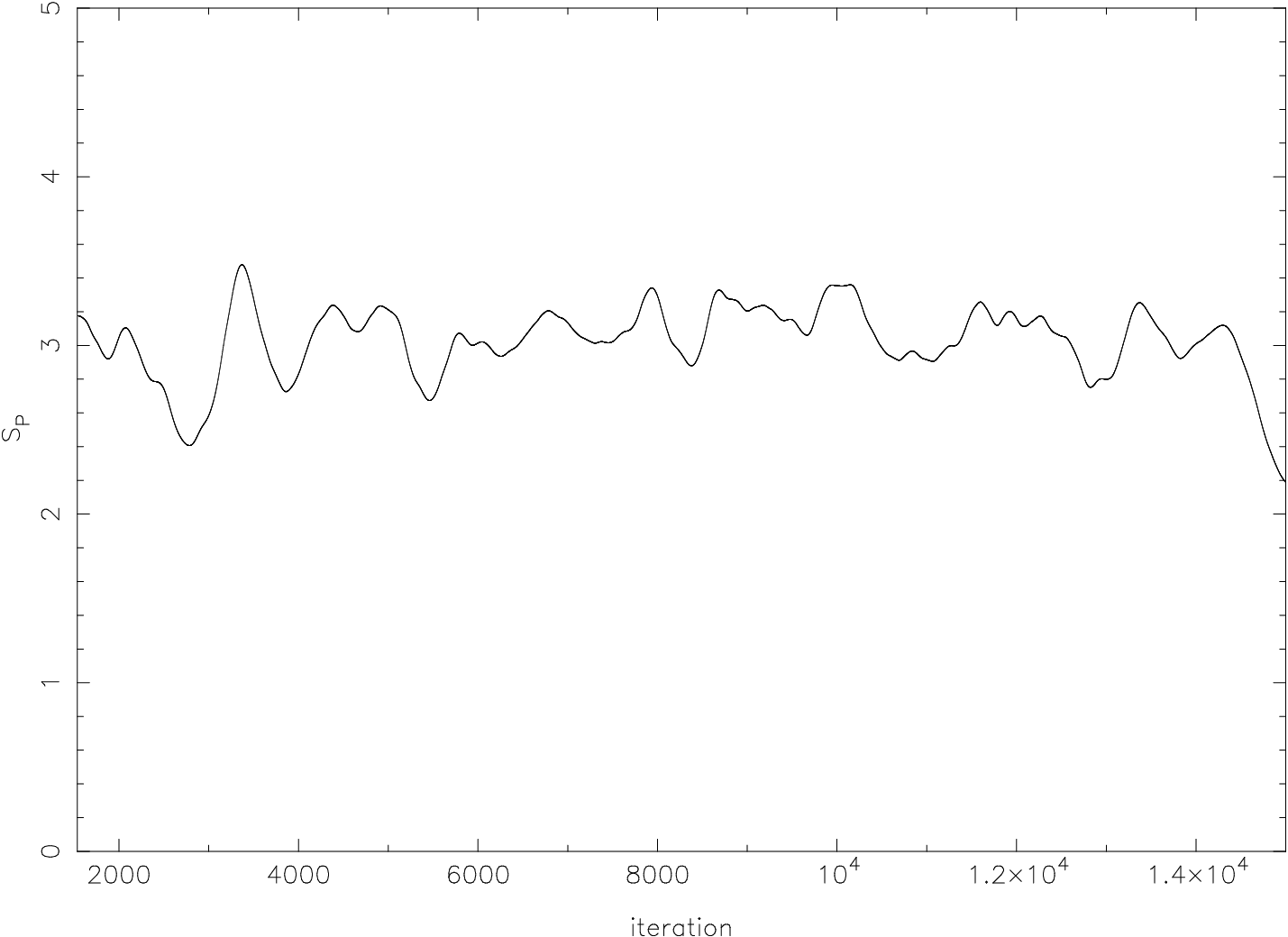}}}
\vskip 0.1cm
\resizebox{\hsize}{!}
{\rotatebox{270}{\includegraphics*{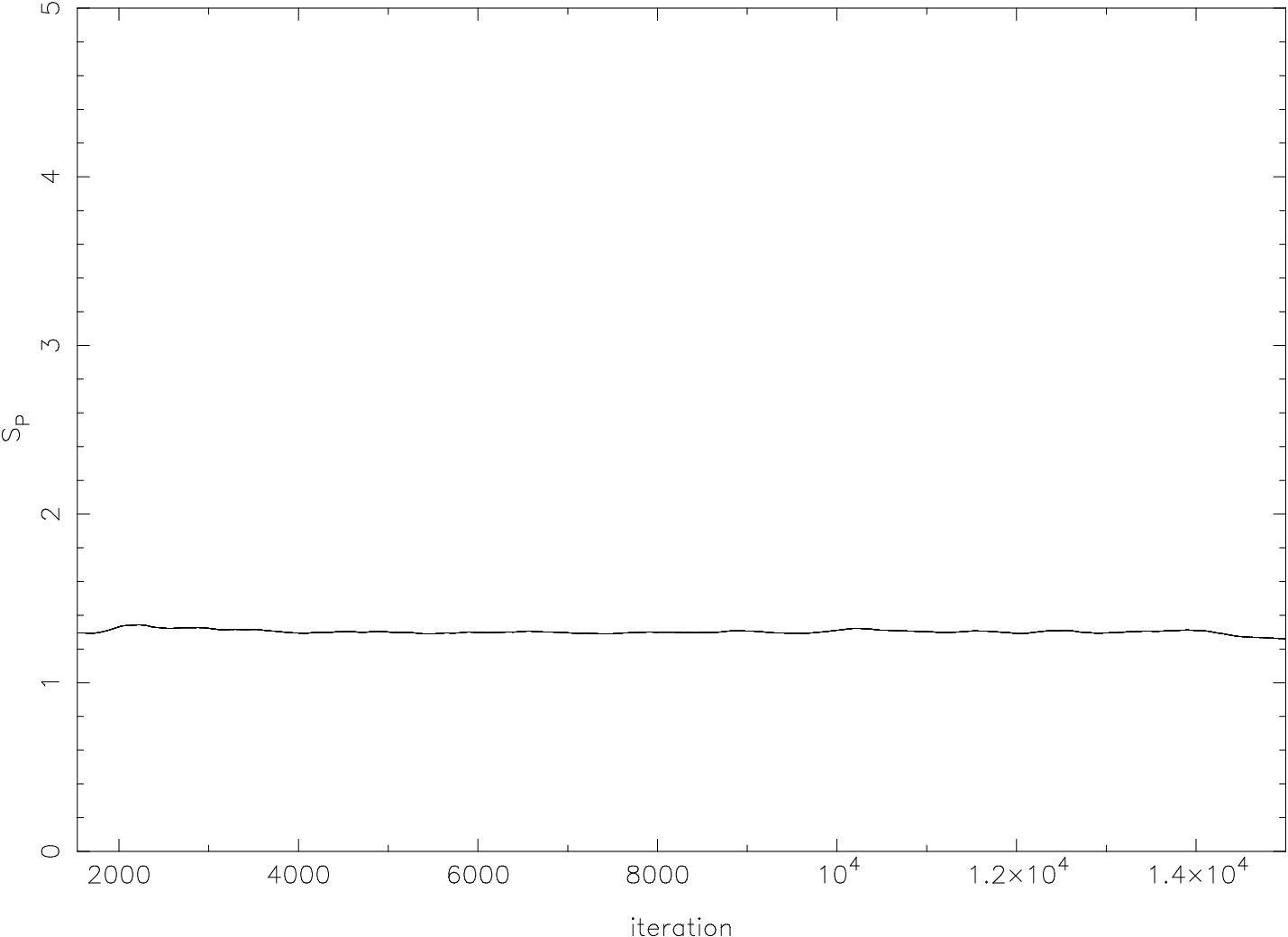}}
\hspace{1cm}
\rotatebox{270}{\includegraphics*{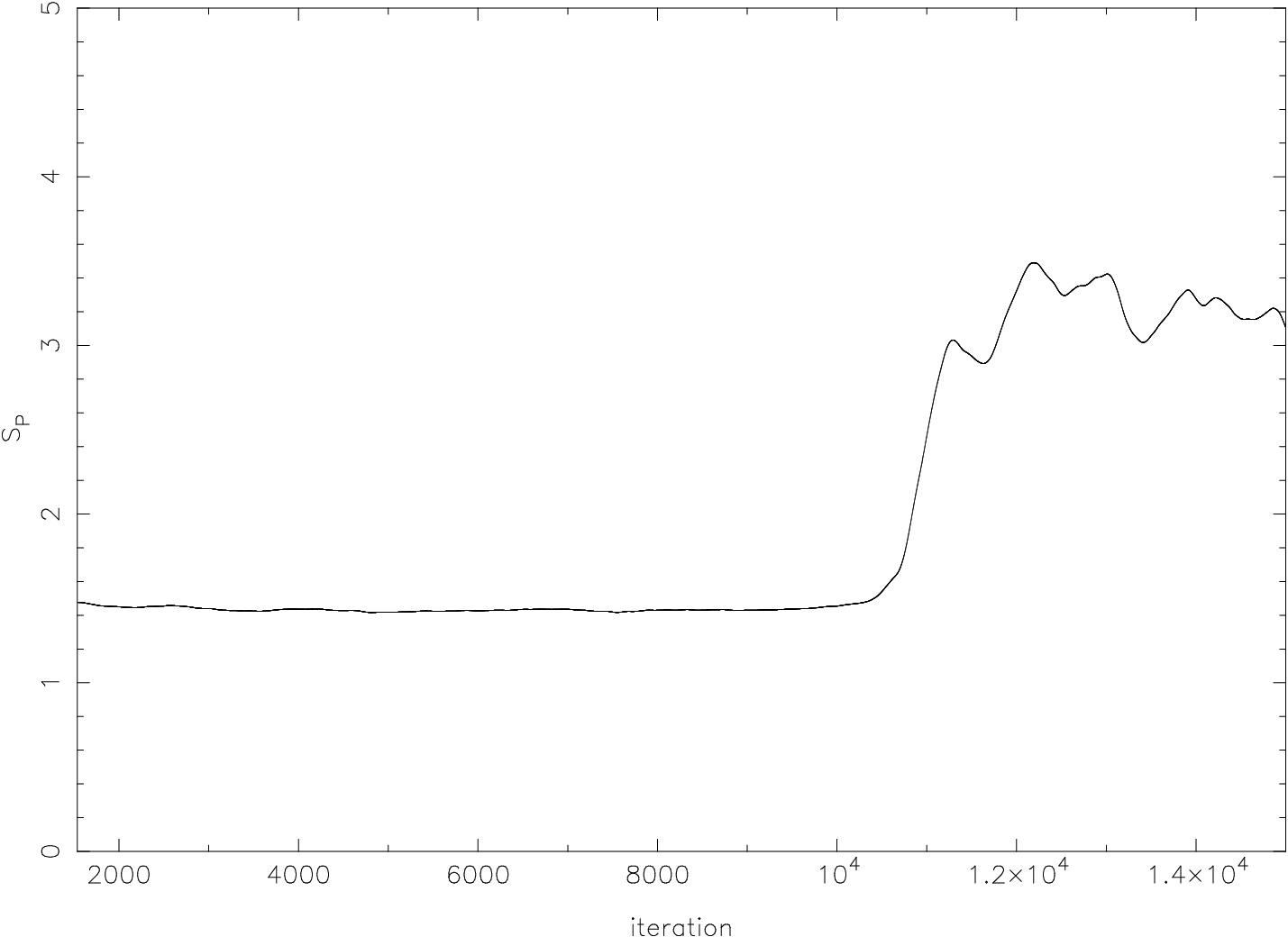}}}
\caption{The evolution of the total power, $S_P(i)$, of the spectra associated
  to the four orbits of figure \ref{fig:chi2}.}  
\label{fig:Sp}
\end{figure*}

The answer to the first question is easy. The two spectra we compare for the
$\chi^2$-analysis, differ significally, not only on the frequencies 
but also on the total power,
\begin{equation}
S_P(i)=\sum_{j=0}^M{P_j(i)^{\prime}} \: ,
\label{eq:Sp}
\end{equation}
of each spectrum. The orbit migrates from a region of weak (i.e., not visually
observable) chaos to a region with strong chaotic behavior. Figure
\ref{fig:Sp} shows the evolution of the total power spectra, $S_P(i)$,
corresponding to the four orbits associated with the $\chi^2(i)$ values of
figure \ref{fig:chi2}. 

In an effort to answer the second question, we (further) ask ourselves: 
{\it "What our results would be, if we compared two power spectra yielded from
  datasets that differ only a few (i.e., a constant number, $n$, of)
  iterations appart?"}  

To do so, instead of taking the first data set as corresponding to the
beginning of the orbit, i.e., to originate from $i=1$, we consider that it
ends at the iteration $i-n$. In this case, the (two) power spectra
correponding to the iteration $i$ are 
\begin{eqnarray}
P_j^{\prime} (i) & = & {\rm PSOD} \left[ q \left ( i-N_S+1 \right ),~...,~q(i) \right ] \nonumber \\
P_j(i) & = & {\rm PSOD} \left [ q \left ( i-N_S+1-n \right ),~...,~q(i-n) \right ]
\label{eq:newPj}
\end{eqnarray}
and their $\chi^2$ likelyhood, which is our new Power Spectrum Indicator
(PSI), to be (from now on) denoted by $\psi^2$, is given by   
\begin{equation}
\psi^2(i) \equiv {\sum_{j=0}^M{\frac{\left ( P_j^{\prime}(i) - P_j(i) \right )^2}{P_j^{\prime}(i) + P_j(i)}}} = \chi^2 (i).
\label{eq:psi2i}
\end{equation}

\begin{figure*}
\resizebox{\hsize}{!}
{\rotatebox{270}{\includegraphics*{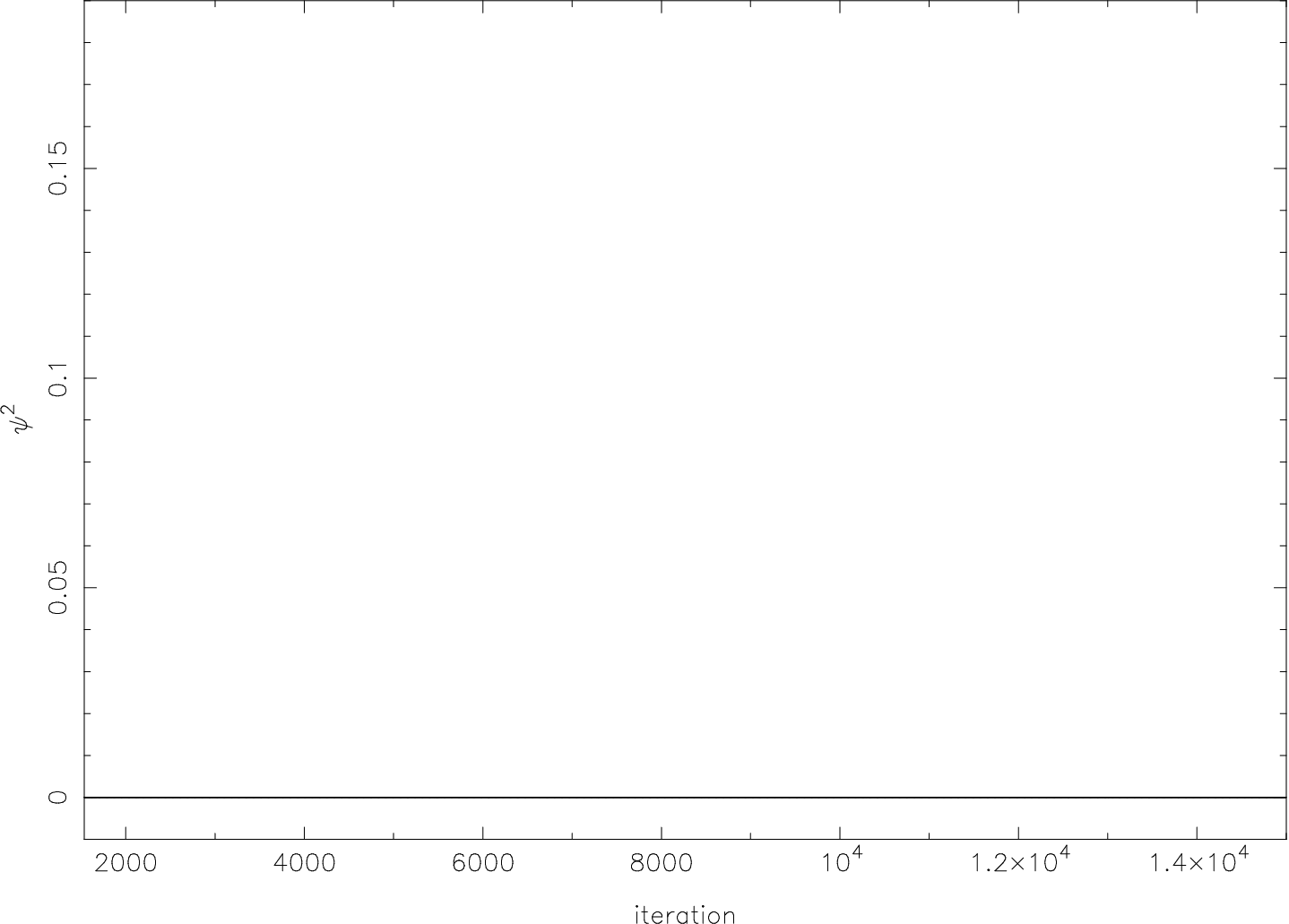}}
\hspace{1cm}
\rotatebox{270}{\includegraphics*{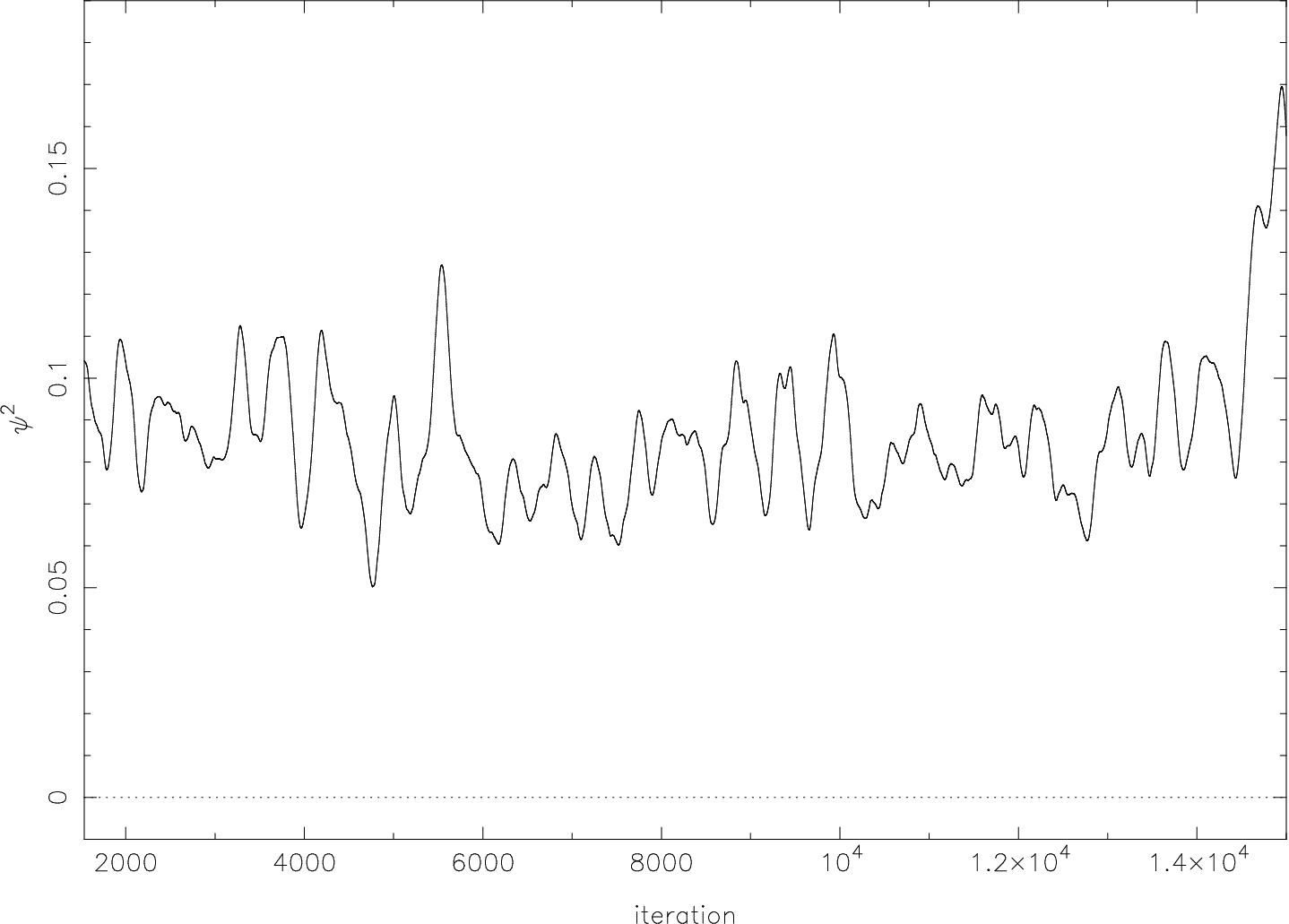}}}
\vskip 0.1cm
\resizebox{\hsize}{!}
{\rotatebox{270}{\includegraphics*{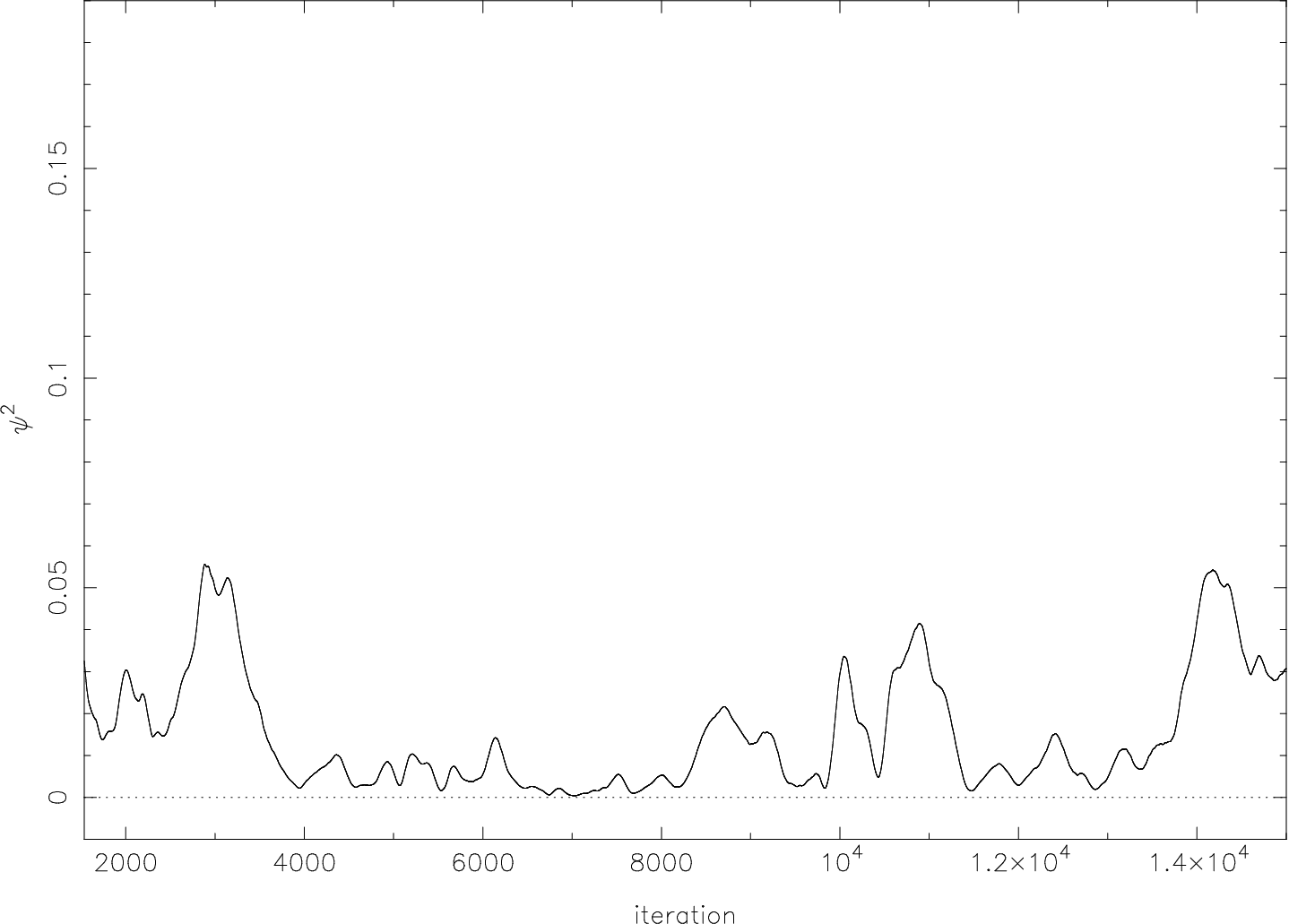}}
\hspace{1cm}
\rotatebox{270}{\includegraphics*{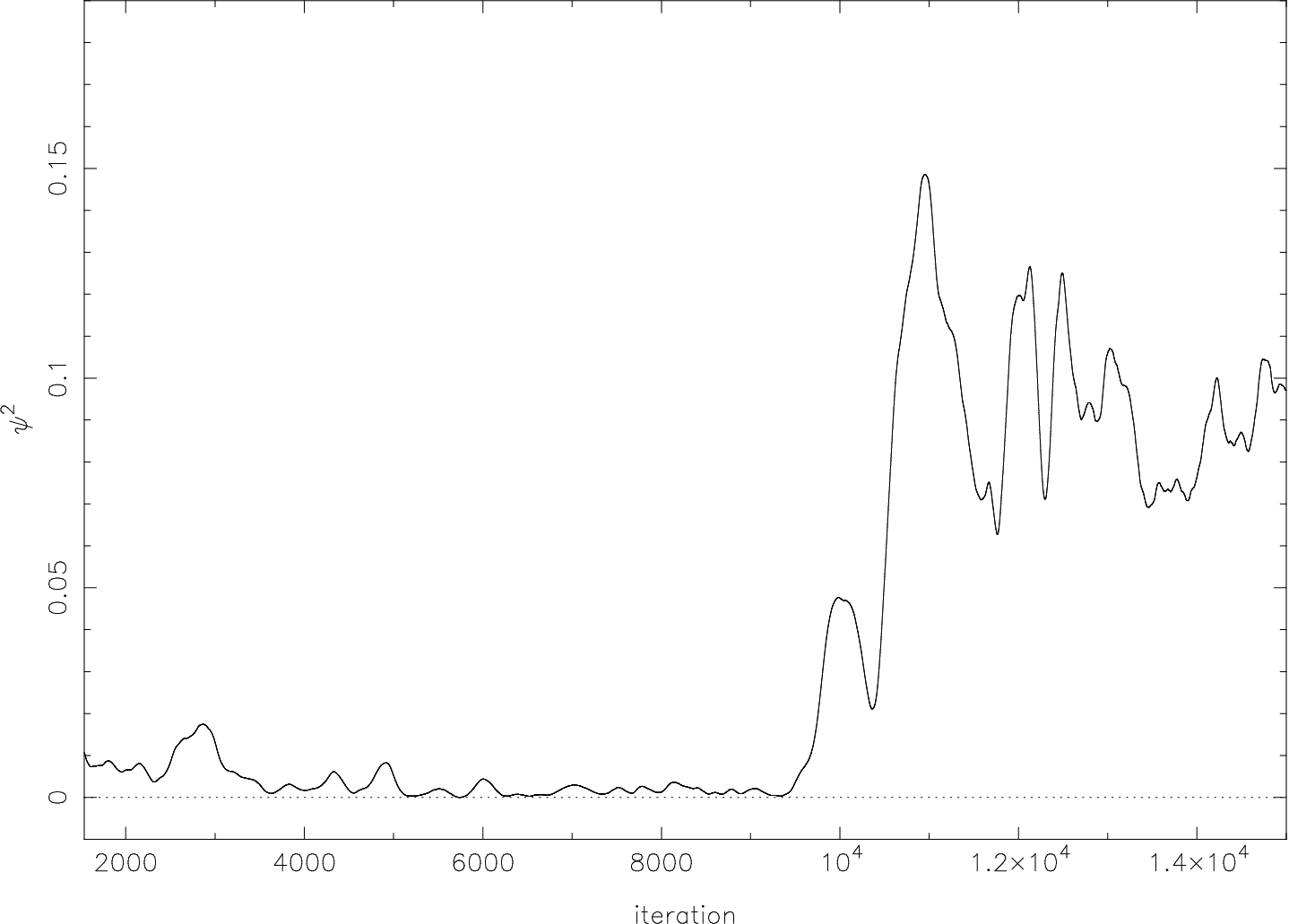}}}
\caption{The evolution of $\psi^2(i)$ associated to the four orbits depicted
  in figure 1. The top left frame corresponds to the ordered orbit,
  originating at $J_0 = \pi $, $\theta_0 = 1.5~\pi$. Top right: The first
  chaotic orbit, originating at $J_0 = 1.3~\pi$, $\theta_0 = 1.5~\pi$. Bottom
  left: The second chaotic orbit, originating at $J_0 = 1.1998~\pi $,
  $\theta_0 = 1.49~\pi$. Bottom right: The power spectrum indicator associated
  to the {\it "sticky"} orbit, originating at $J_0 = ~\pi $, $\theta_0 =
  1.538~\pi$. In each and every one of these cases, the datasets involved
  differ with each other by $n = 256$ iterations.}  
\label{fig:psi2}
\end{figure*}

Figure \ref{fig:psi2} presents the evolution of our new indicator, $\psi^2$,
as a function of the iteration number $i$, regarding the four orbits of figure
1. The two spectra are calculated from datasets that are $n = 256$ iterations
apart. As expected, for the ordered orbit (upper left frame) the value of
$\psi^2$ is always zero. On the contrary, the two chaotic orbits (upper right
and lower left) exhibit a clearly non-zero value of $\psi^2$. This value is by
no means constant, because, at different values of the iteration number, $i$,
the orbit visits different areas of the chaotic region. Notice that, the
values of $\psi^2$ corresponding to the first chaotic orbit (upper right
frame), are much higher than those of the second one (lower left frame),
indicating that the former orbit is in a region of {\it "strong"} chaos, as
compared to the {\it "weaker"} chaotic region in which the latter orbit lies
up.  

Finally, as regards the power spectrum indicator, $\psi^2$, associated with
the {\it "sticky"} orbit (lower right frame of figure \ref{fig:psi2}), it
initially admits a low, but clearly non-zero value, indicating that it rests
in a {\it "weak"} chaotic region. However, after a lot of iterations, the
orbit migrates from the {\it "weak"} chaotic region to a region with 
{\it "stronger chaos"}, analogous to the region of the first chaotic
orbit. Accordingly, the value of $\psi^2$ climbs to higher levels, exhibitting
a behavior similar to that of the first chaotic orbit. 

\begin{figure*}
\resizebox{\hsize}{!}
{\rotatebox{270}{\includegraphics*{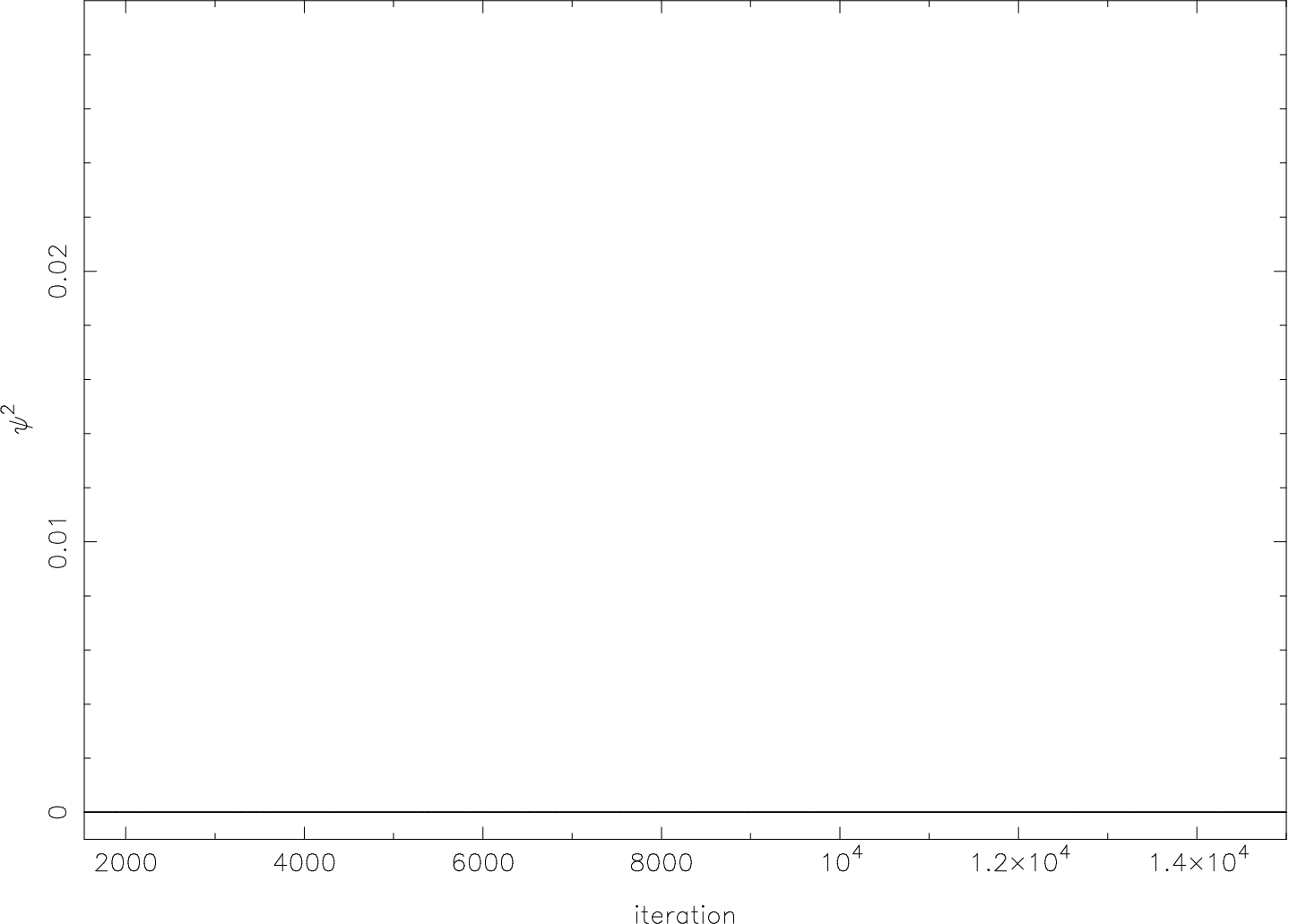}}
\hspace{1cm}
\rotatebox{270}{\includegraphics*{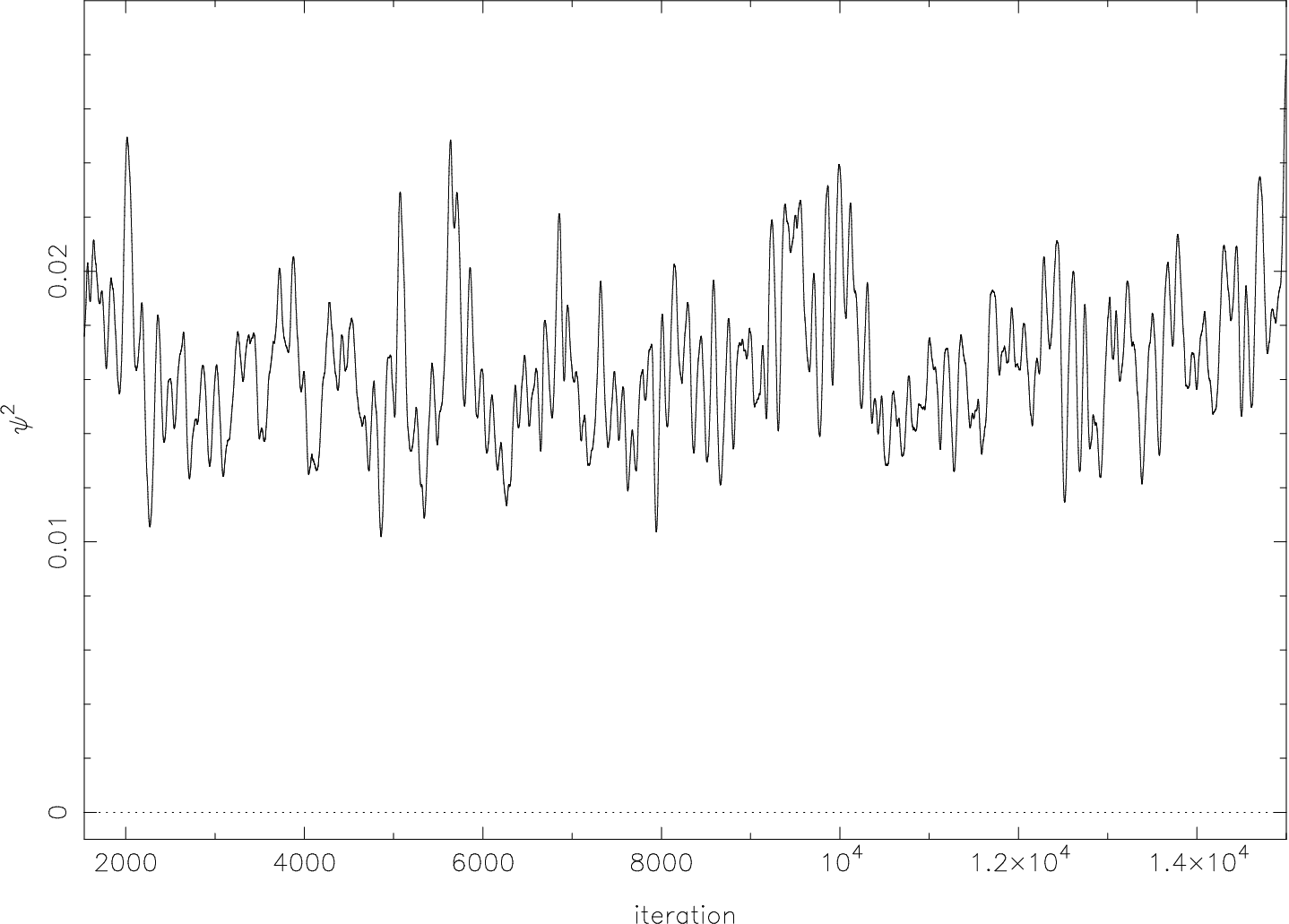}}}
\vskip 0.1cm
\resizebox{\hsize}{!}
{\rotatebox{270}{\includegraphics*{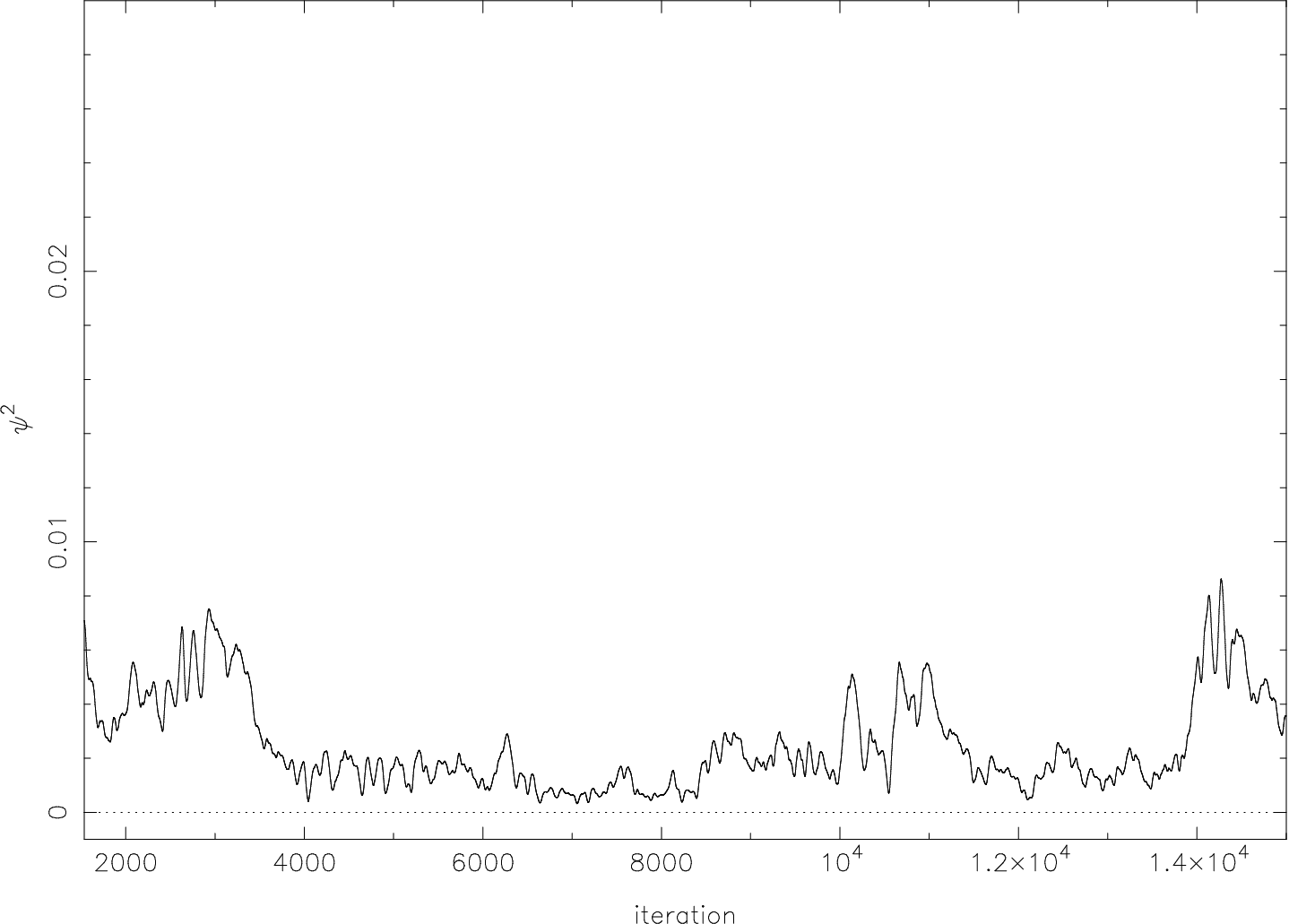}}
\hspace{1cm}
\rotatebox{270}{\includegraphics*{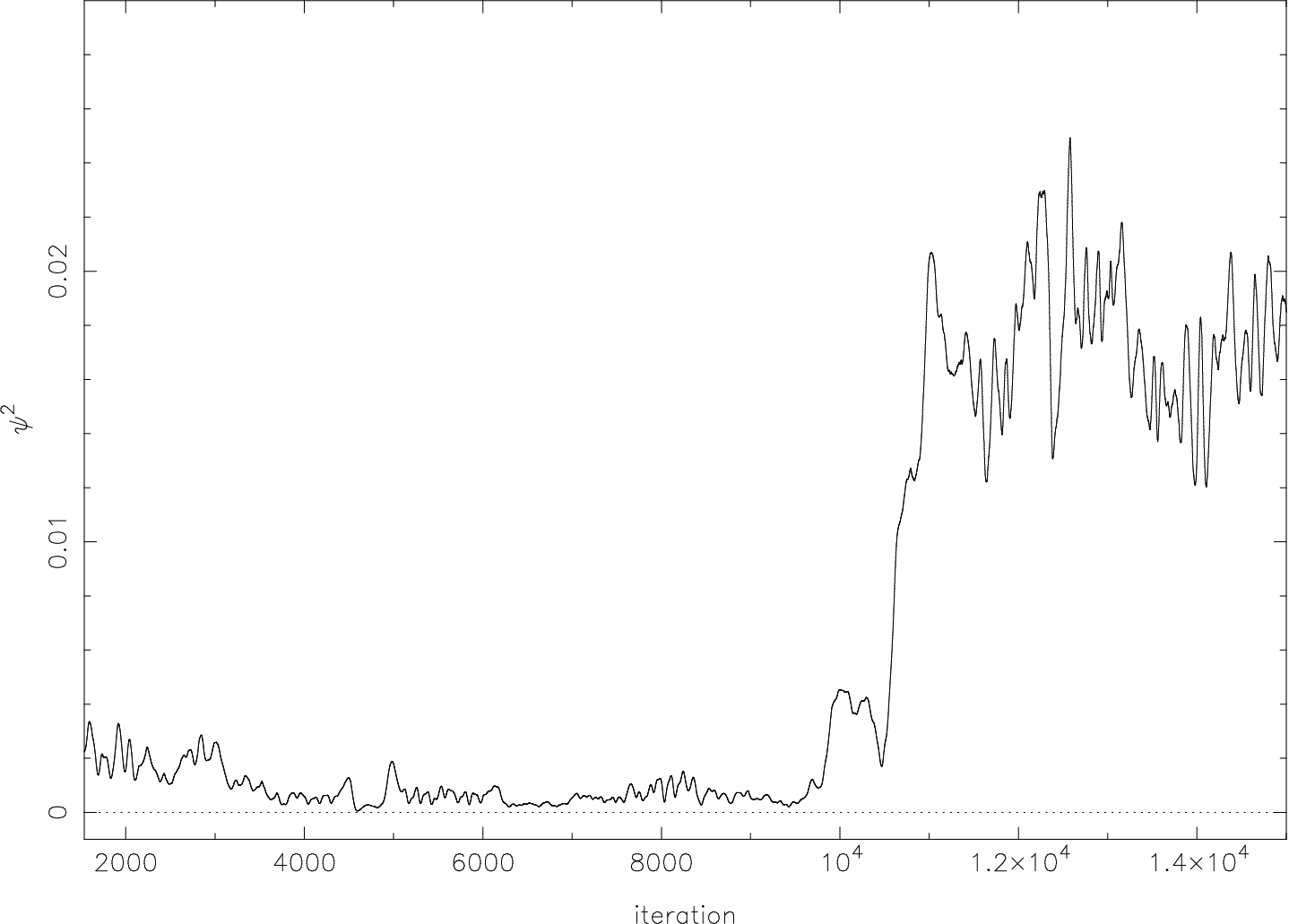}}}
\caption{Same as figure \ref{fig:psi2}, but for $n = 64$.} 
\label{fig:psi2a}
\end{figure*}

At this point, we need to stress that, the potential values of $\psi^2$ depend
on the seperation $n$ of the two datasets. The lower the value of $n$, the
closer the two datasets are, hence, their $\psi^2$ will be lower. Note that
for $n < N_s$ the two datadets have $N_s - n$ common data. Nevertheless, even
for $n = 64$ ($N_s = 3 \times 256$) the PSI method gives quite accurate
results, as it can be readily seen from figure \ref{fig:psi2a}. 

\begin{figure*}
\resizebox{\hsize}{!}
{\rotatebox{270}{\includegraphics*{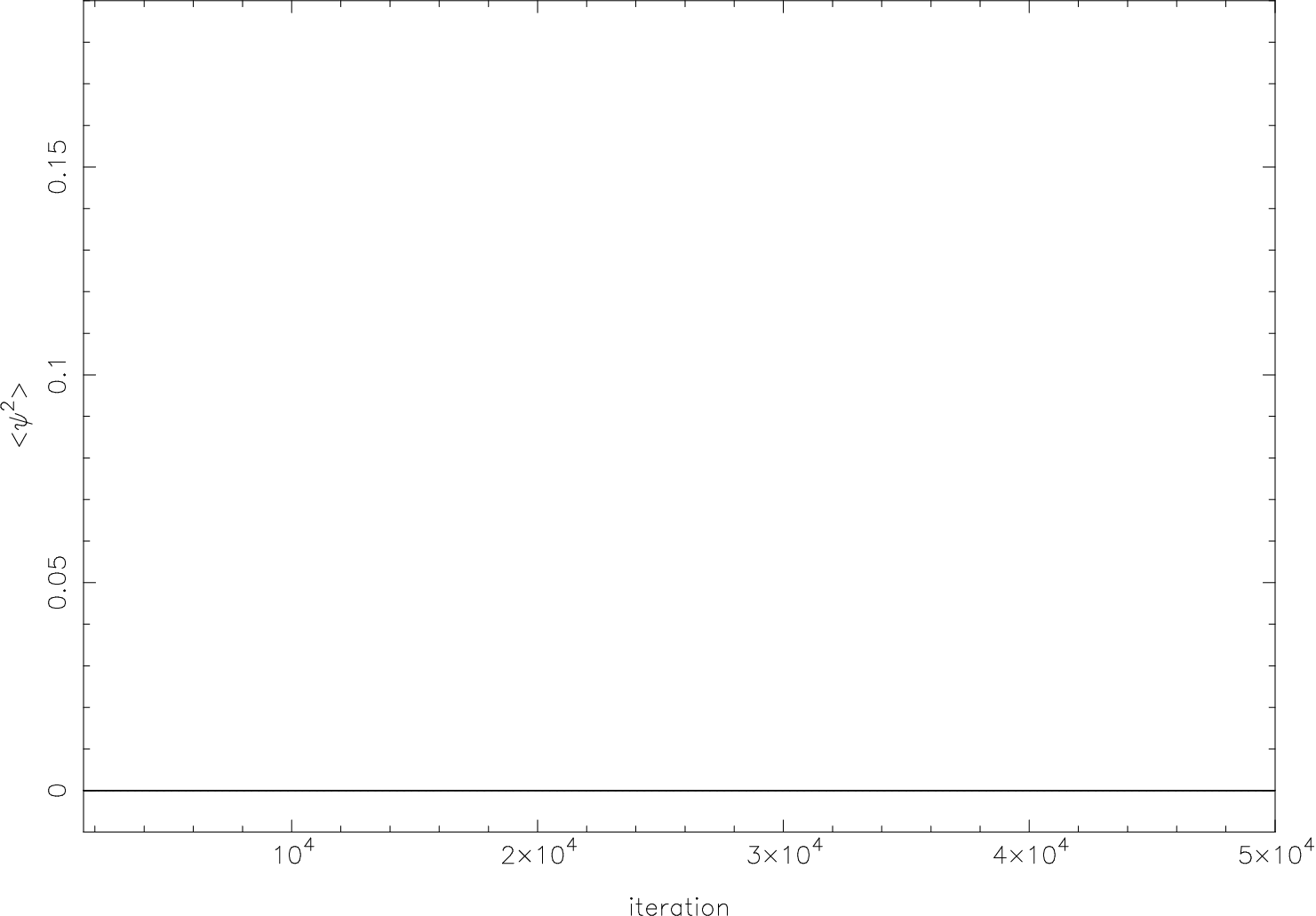}}
\hspace{1cm}
\rotatebox{270}{\includegraphics*{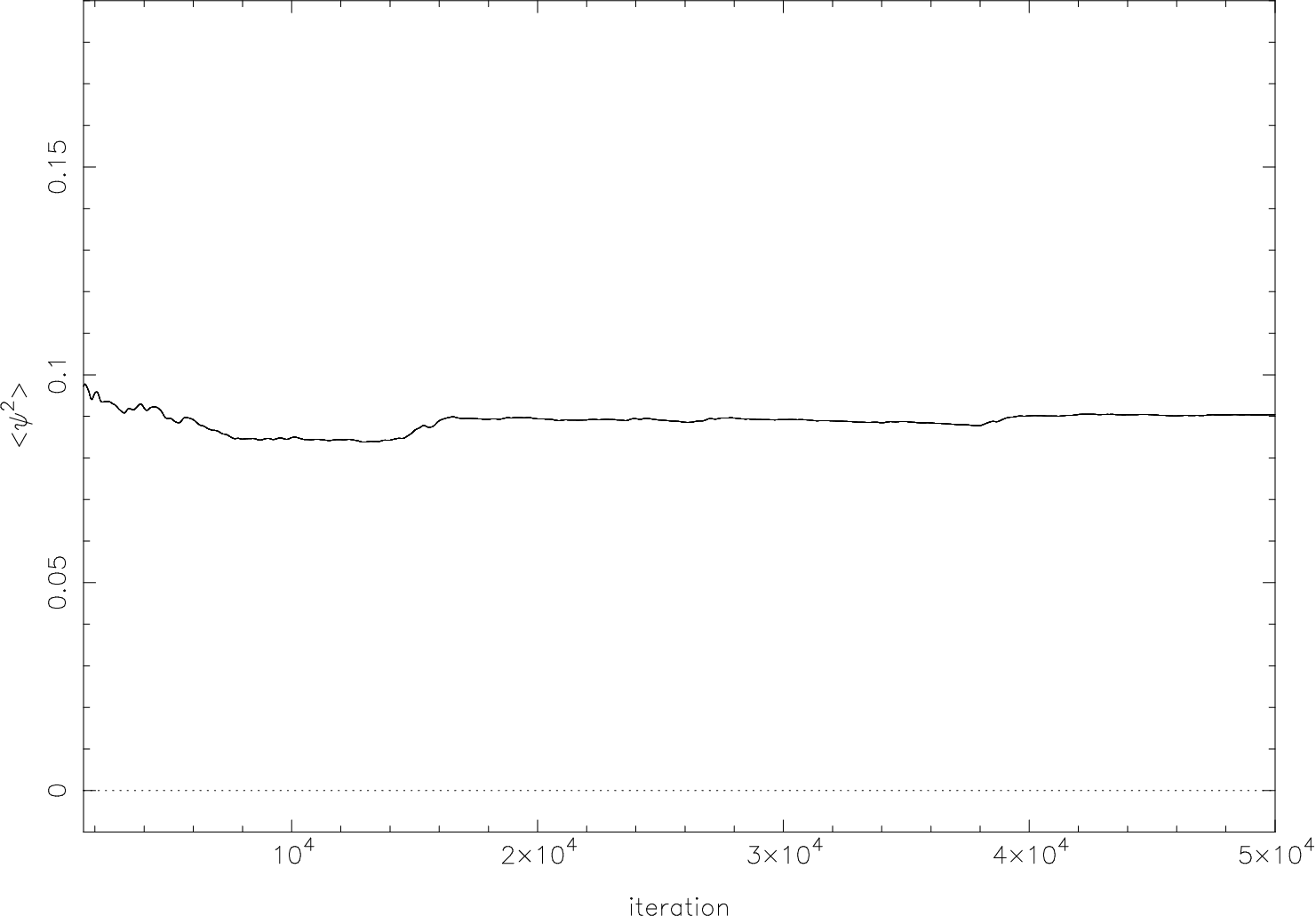}}}
\vskip 0.1cm
\resizebox{\hsize}{!}
{\rotatebox{270}{\includegraphics*{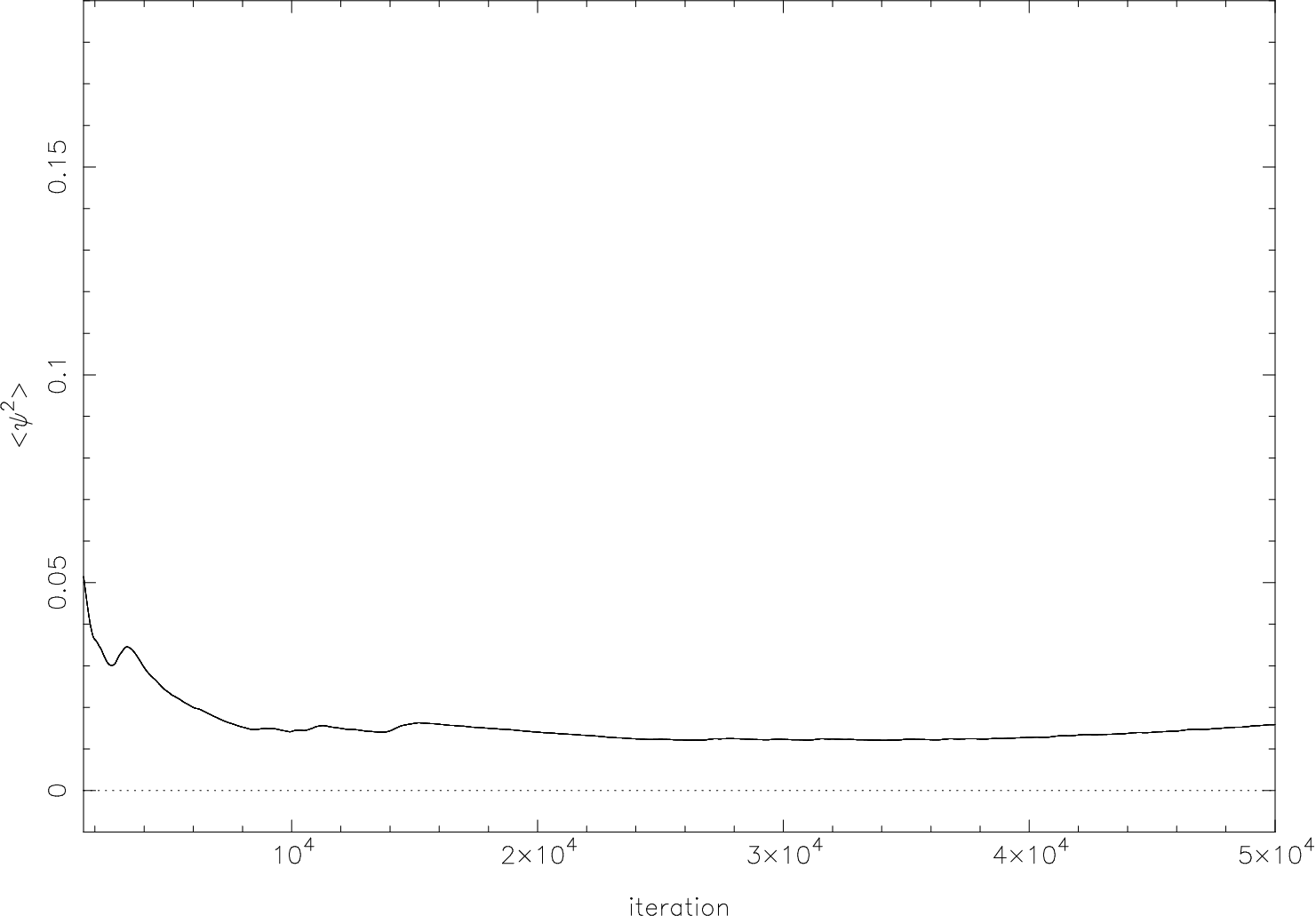}}
\hspace{1cm}
\rotatebox{270}{\includegraphics*{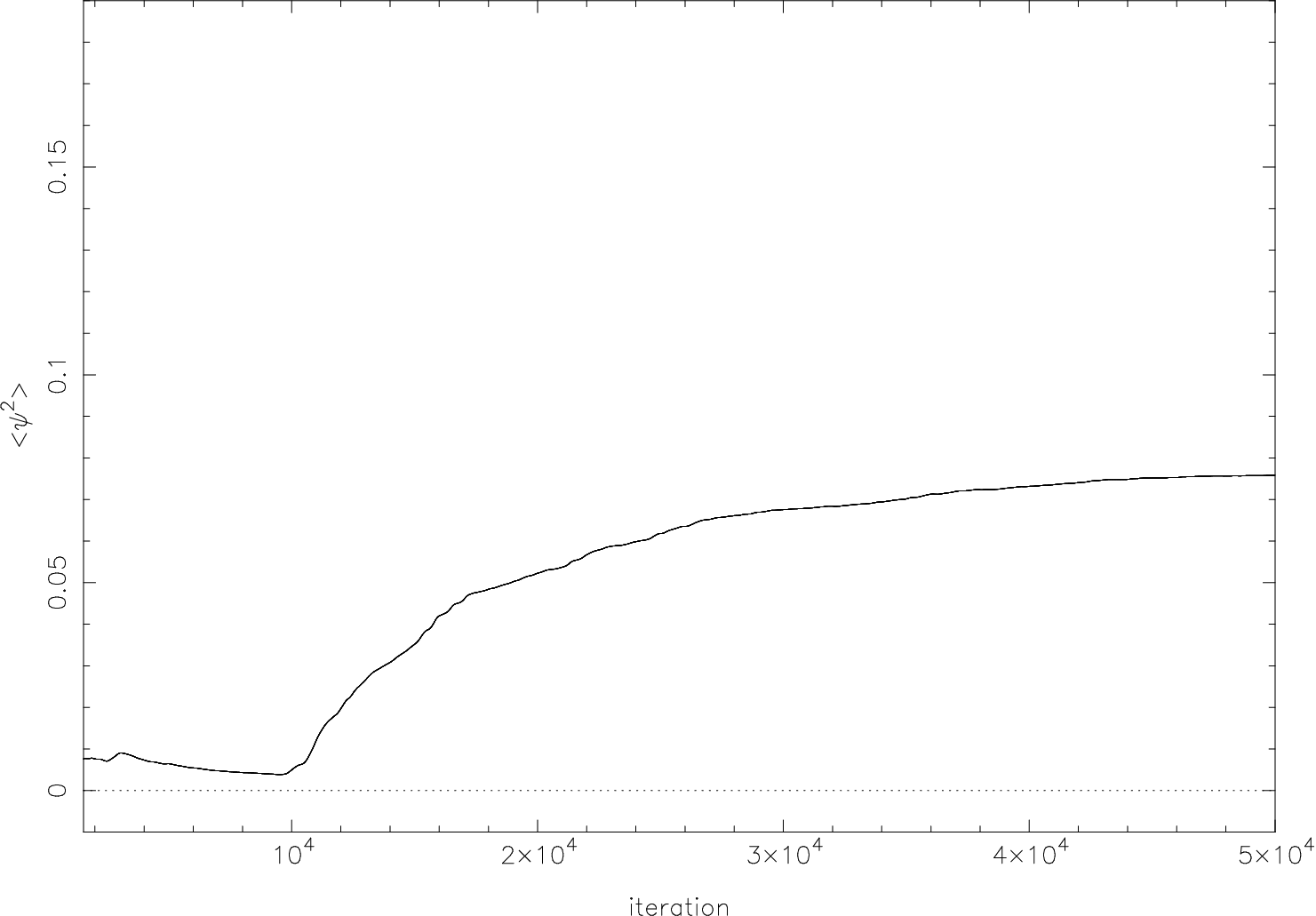}}}
\caption{The evolution of the global indicator $\left \langle \psi^2 \right
  \rangle (i)$, for $n = 256$, as regards the four orbits presented in figure
  1. Top left, the ordered orbit originating at $J_0 = \pi $, $\theta_0 =
  1.5~\pi$; Top right, the chaotic orbit originating at $J_0 = 1.3~\pi $,
  $\theta_0 = 1.5~\pi$; Botton left, the second chaotic orbit originating at
  $J_0 = 1.1998~\pi $, $\theta_0 = 1.49~\pi$; Bottom right, the {\it "sticky"}
  orbit originating at $J_0 = ~\pi $, $\theta_0 = 1.538~\pi$.}  
\label{fig:psi2m}
\end{figure*}

As mentioned above, the value of $\psi^2(i)$ give us a local (i.e., in the
region currently visited by the orbit) indication of how strong the chaos may
be. In oder to attain a global indicator (i.e., one that will cover the whole
area visited by an orbit), we should consider the average value of $\psi^2$,
defined as 
\begin{equation}
\left \langle \psi^2 \right \rangle (i) = \frac{1}{i-N_s-n+1}{\sum_{j=N_s+n}^i \psi^2(j)}.
\label{eq:psi2ai}
\end{equation}
Figure \ref{fig:psi2m} shows the evolution of $\left \langle \psi^2 \right
\rangle$ for the four orbits considered throughout this article. Their
$\psi^2$ indicators are computed upon the use of a dataset seperation of $n =
256$ iterations.  

\section{Conclusions}

In the present paper we propose a new tool, to be called Power Spectrum
Indicator (PSI), or $\psi^2$, that enables us to determine, as early as
posible, the chaotic nature of orbits in dynamical systems. This new method is
based on the method of Vozikis et al. (2000), i.e., on the frequency analysis
of a data series constucted by recording the logarithm of the amplification
factor of the deviation vector of nearby orbits. For this reason, two datasets
are recorded and the $\chi^2$-likelyhood of their power spectra is computed.  
 
Ordered orbits have always the same power spectrum, so their $\psi^2 \equiv
\chi^2$ acquires an ever-zero value. On the contrary, a chaotic orbit has a
power spectrum that varies with time (equivalently, with the number of
iterations). Therefore, chaotic orbits always exhibit a non-zero $\psi^2$
value, hence, by calculating the $\psi^2$ of an orbit, we can easily decide if
this is chaotic ot not. Even for {\it "sticky"} orbits, the PSI method
is very effective in the early detection of chaos. Eventually, the global
behavior of the $\psi^2$ indicator can provide information (also) on the
intense of the chaotic behavior, i.e., on how {\it "strong"} or {\it "weak"}
the associated chaos may be.  

However, we need to stress that, the aforementioned results refer to the case
where the system under study is a $2D$ mapping. Further investigation is
needed, if the $\psi^2$ method is to be implemented also in Hamiltonian
flows. As stated in the discussion of Vozikis et al. (2000), the difference
between maps and flows is that, in the case of flows, a detailed analysis
concerning the selection of the renormalization time is needed, since it may
significantly affect the corresponding results.

\ack{Financial support by the Research Committee of the Technological
Education Institute of Central Macedonia at Serres, Greece, under grant
SAT/CE/201217-aa/01, is gratefully acknowledged.}

\References

\item[] Aubry S and Abramovici G 1990 {\it Physica D} {\bf 43} 199
\item[] Benettin G, Galgani L and Strelcyn J M 1976 \PR A  {\bf 14} 2338
\item[] Contopoulos G 1966 {\it Les Nouvelles M\'ethodes de la Dynamique
Stellaire} ed F Nahon and M H\'enon (Paris: CNRS)
\item[] Contopoulos G and Voglis N 1997 {\it Astron. Astroph.} {\bf 317} 73
\item[] H\'enon M and Heiles C 1964 \AJ {\bf 69} 73
\item[] Froeschl\'e C, 1984 {\it Cel. Mech.} {\bf 34} 95 
\item[] Froeschl\'e C, Froeschl\'e Ch and Lohinger E 1993 {\it
  Cel. Mech. Dyn. Astron.} {\bf 51} 135
\item[] Froeschl\'{e} C, Lega E and Gonzi R 1997, {\it Cel. Mech. Dyn. Astron.} 
{\bf 67} 41
\item[] Gelfreich V G 1999 {\it Commun. Math. Phys.} {\bf 201} 155
\item[] Ichikawa Y H, Kamimura T and Hatori T 1987 {\it Physica D} {\bf 29} 247
\item[] Karanis G and Vozikis Ch 2008 {\it Astron. Nachr.} {\bf 320} 403
\item[] Laskar J 1993 {\it Physica D} {\bf 67} 257
\item[] Laskar J, Froeschl\'e C and Celleti A 1992 {\it Physica D} {\bf 56}
  253
\item[] Lazutkin V F 2005 {\it J. Math. Science} {\bf 128} 2687
\item[] Lichtenberg A J and Lieberman M A 1983 {\it Regular and stochastic
  motion} (New York: Springer) pp. 77-84) 
\item[] Press W H, Teukolsky S A, Vetterling W T and Flannery B P 1992
{\it Numerical Recipes in Fortran -- The Art of Scientific Computing 2nd
edn} (Cambridge: Cambridge University Press)
\item[] Skokos Ch 2001 \JPA {\bf 34} 10029
\item[] Skokos Ch, Antonopoulos C, Bountis A and Vrahatis M N 2003 {\it
  Prog. Theor. Phys. Suppl.} {\bf 150} 439 
\item[] Skokos Ch, Antonopoulos C, Bountis A and Vrahatis M N 2004 \JPA {\bf
  37} 6269
\item[] Skokos Ch, Bountis A and Antonopoulos C G 2007 {\it Physica D} {\bf
  231} 30  
\item[] Skokos Ch and Manos T 2016 {\it Chaos Detection and Predictability
}({\it Lecture Notes in Physics vol 915}) ed  Skokos Ch, Gottwald G and Laskar
  J  (Berlin, Heidelberg : Springer)
\item[] Voglis N and Contopoulos G 1994, \JPA {\bf 27} 4899
\item[] Voyatzis G and Ichtiaroglou S 1992 \JPA {\bf 25} 5931
\item[] Vozikis Ch L, Varvoglis H and Tsiganis K 2000 {\it Astron. Astrophys.}
  {\bf 359}  386 

\endrefs

\end{document}